\documentclass[twocolumn,superscriptaddress,aps,jcp,amsmath,amssymb]{revtex4-1}

\usepackage[latin9]{inputenc}
\usepackage{amsthm}
\usepackage{amsmath}
\usepackage{amssymb}
\usepackage{graphicx}
\usepackage{esint}
\usepackage{hyperref}
\usepackage{paralist}
\usepackage{xcolor} 
\usepackage{epstopdf}

\graphicspath{{Figures/}}

\def\C{{\mathbb C}}% complex numbers
\def\R{{\mathbb R}}% real numbers
\def\RR{{\mathbb{R}}}
\def\CC{{\mathbb{C}}}
\def\ge{\geqslant}%greaterorequal

\newcommand{\wh}{\widehat}
\newcommand{\wt}{\widetilde}
\newcommand{\ud}{\,\mathrm{d}}

\IfFileExists{mathabx.sty}%
  {\DeclareFontFamily{U}{mathx}{\hyphenchar\font45}%
   \DeclareFontShape{U}{mathx}{m}{n}{<->mathx10}{}%
   \DeclareSymbolFont{mathx}{U}{mathx}{m}{n}%
   \DeclareFontSubstitution{U}{mathx}{m}{n}%
   \DeclareMathAccent{\widebar}{0}{mathx}{"73}%
}{%
  \PackageWarning{mathabx}{%
    Package mathabx not available, therefore\MessageBreak substituting
    widebar with overline\MessageBreak }%
  \newcommand{\widebar}[1]{\overline{#1}}%
}
\newcommand{\wb}[1]{\widebar{#1}}

\newcommand{\bd}{\boldsymbol}

\theoremstyle{plain}

\theoremstyle{definition}

\newtheorem*{remark*}{Remark}

\newcommand{\mc}[1]{\mathcal{#1}}

\newcommand{\abs}[1]{\lvert#1\rvert}
\newcommand{\Abs}[1]{\left\lvert#1\right\rvert}

\newcommand{\norm}[1]{\lVert#1\rVert}

\newcommand{\bra}[1]{\langle#1\rvert}

\newcommand{\ket}[1]{\lvert#1\rangle}

\newcommand{\dps}{\displaystyle}

\DeclareMathOperator{\tr}{Tr}

\newcommand{\Var}{\mathrm{Var}}

\begin{document}

\title{Accelerated sampling by infinite swapping of path integral
  molecular dynamics with surface hopping}

\author{Jianfeng Lu} \email{jianfeng@math.duke.edu}
\affiliation{Department of Mathematics, Department of Physics and
  Department of Chemistry, Duke University, Durham NC 27708, USA}
\author{Zhennan Zhou} \affiliation{Beijing International Center of
  Mathematical Research, Peking University, Beijing 100871,
  P.R. China}

\begin{abstract}
  To accelerate the thermal equilibrium sampling of multi-level
  quantum systems, the infinite swapping limit of a recently proposed
  multi-level ring polymer representation is investigated. In the
  infinite swapping limit, the ring polymer evolves according to an
  averaged Hamiltonian with respect to all possible surface index
  configurations of the ring polymer{, thus connects the surface
    hopping approach to the mean-field path integral molecular
    dynamics}. A multiscale integrator for the infinite swapping limit
  is also proposed to enable {efficient} sampling based
  on the limiting dynamics. Numerical results demonstrate the huge improvement of
  sampling efficiency of the infinite swapping compared with the
  direct simulation of path integral molecular dynamics with surface
  hopping.
\end{abstract}

\maketitle

%%%%%%%%%%%%%%%%%%%%%%%%%%%%%%%%%%%%%%%%%%%%

\section{Introduction}

This work aims at designing efficient sampling methods of thermal
averages of multi-level quantum systems
\begin{equation}\label{eq:aveA} 
  \langle\wh{A}\rangle = \frac{\tr_{ne}
    [e^{-\beta \wh H} \wh A ] }{\tr_{ne}[e^{-\beta \wh H}]},
\end{equation} 
where $\wh{H}$ is a multi-level Hamiltonian operator, $\wh{A}$ is a
matrix observable, and $\beta$ is the inverse temperature. The
multi-level quantum systems arise when the non-adiabatic effect
between different energy surfaces of electronic states cannot be
neglected, see e.g., the review articles \cite{Makri:99,
  StockThoss:05, Kapral:06}.

The ring polymer representation, originally proposed in
\cite{Feynman:72}, is an effective way based on path integral to map
the quantum thermal average problem to a classical thermal average in
an extended phase space. The representation serves as the foundation
for various methods, including the path integral Monte Carlo
\cite{ChandlerWolynes:81, BerneThirumalai:86} and path integral
molecular dynamics \cite{MarklandManolopoulos:08,
  CeriottiParrinelloMarklandManolopoulos:10} sampling techniques.  

For the multi-level quantum system as in \eqref{eq:aveA}, the ring
polymer representation can be extended so that each bead in the
polymer is associated with a surface index taken into account the
multiple levels {
  \cite{SchmidtTully2007,ShushkovLiTully:12,LuZhouPIMDSH,DukeAnanth2016,ShakibHuo2017}.}
Based on the multi-level ring polymer representation, in our previous
work \cite{LuZhouPIMDSH}, a path integral molecular dynamics with
surface hopping (PIMD-SH) method is proposed for thermal average
calculations, where the discrete electronic state is sampled by a
consistent surface hopping algorithm coupled with Hamiltonian dynamics
of the position and momentum with Langevin thermostat.  Such surface
hopping type dynamics for thermal (imaginary time) sampling can be
naturally combined with real time surface hopping dynamics
\cite{Tully:90, HammesSchifferTully:94, Tully:98, Kapral:06,
  ShenviRoyTully:09, Barbatti:11, SubotnikShenvi:11, Subotnik:16,
  FGASH, FGASH2}.  {It is worth emphasizing that, different from
  \cite{ShushkovLiTully:12,ShakibHuo2017} where the hopping rate were
  empirically imposed based on the real time surface hopping dynamics,
  the hopping process in our previous work \cite{LuZhouPIMDSH}
  satisfies the detailed balance condition and thus the PIMD-SH
  samples the \emph{exact} equilibrium distribution of the ring
  polymer configuration, up to numerical discretization error.  Other
  than the surface hopping approach for which the bead degree of
  freedom is augmented with a discrete surface index, one could also
  use the mapping variable approach \cite{MeyerMiller:79,
    StockThoss:97} to extend the conventional ring polymer
  representation to multi-level systems; see the review article
  \cite{StockThoss:05} and more recent developments in
  \cite{AnanthMiller:10, RichardsonThoss:13,
    Ananth:13,MenzeleevBellMiller:14,CottonMiller:15,
    HeleAnanth:16,JianLiu:16}} {Let us also mention the related sampling approaches for reduced density matrix for open quantum systems of a system coupled to harmonic bath \cite{MoixZhaoCao2012,MCReichman2017}.}

While the PIMD-SH method has been validated through numerical examples
in \cite{LuZhouPIMDSH}, it is observed that sampling off-diagonal
elements of the observable, associated with the coupling between
different energy surfaces, is more challenging. This is due to the
fact that in the ring polymer representation, only consecutive beads
living on different energy surfaces (referred to as a kink) can make
major contribution to the average; while these kinks are rare to form
in the PIMD-SH dynamics since such configuration has higher energy
compared to configurations without kinks.

As noted in \cite{LuZhouPIMDSH}, it is possible to increase the
hopping frequency $\eta$ so that the ring polymer develops kinks on a
faster time scale, in fact, it can be proved that the sampling
efficiency increases as the hopping intensity parameter
increases. However, at the same time, a faster hopping dynamics
increases the stiffness of the dynamics and hence makes the numerical
integration more challenging.

To overcome this difficulty, in this work, we will investigate the
infinite swapping limit of the ring polymer representation for
multi-level quantum systems.  When the hopping frequency $\eta \gg 1$,
there are two distinct time scales in the trajectory evolution under
PIMD-SH: the fast scale is characterized by the typical time length of
changes in the surface index (hopping), and the slow scale corresponds
to the Langevin dynamics. Thus, the infinite swapping limit can be
viewed as the homogenization limit of a multiscale dynamical systems
\cite{PSbook,Ebook}.  As we will show, as the hopping intensity
parameter $\eta$ goes to infinity, the averaging leads to an explicit
limiting dynamics of only the slow variables, the position and
momentum of beads in the ring polymer, while the fast variables are
effectively in local equilibrium and can be integrated out.  {We
  show that the PIMD-SH converges to the mean-field ring polymer
  representation \cite{PrezhdoRossky:97,DukeAnanth2016} in the
  infinite swappling limit. This connection facilitates the design of
  numerical integrators for sampling the ring polymer configuration
  for thermal averages.}

To simulate the infinite swapping dynamics, we borrow the idea from
heterogeneous multiscale methods (HMM)
\cite{EEngquist:03,EVE:03,Weinan2007,Ebook} for multiscale dynamics
systems ({see also earlier works of multiple time-stepping
  methods with similar ideas \cite{GearWells84, Tuckerman1990,
    TuckermanBerne1991, TuckermanBerne1991_2, TuckermanBerne1991_3}}),
in particular, the recently developed multiscale integrator for
replica exchange method when the swapping frequency is high
\cite{TangLuAbramsVE, LuVEremd} by one of us.  Our proposed multiscale
integrator effectively samples the ring polymer representation for
$\eta \gg 1$, while avoiding the enumeration of all possible surface
index configurations at every time step. Following the spirit of
\cite{TangLuAbramsVE}, the multiscale integrator consists of a
macrosolver for the Langevin dynamics and a microsolver for the
continuous-time Markov jump process of the surface index, with the two
linked by an estimator passing information from micro- to
macro-solvers. The details are presented in \S\ref{sec:hmm}.  The
multiscale integrator efficiently samples the thermal average based on
the infinite swapping dynamics, especially when the potential
landscape is complicated or the temperature is low, so that it is
necessary to use more beads in the ring polymer representation.

The paper is outlined as follows. In Section \ref{sec:theory}, we
briefly review the ring polymer representation for the thermal
averages for multi-level quantum systems and its direct simulation
method. The derivation of the infinite swapping limit is then
presented in detail, with discussions on its efficiency compared to
the PIMD-SH method. We discuss the numerical algorithms in simulating
the PIMD-SH and its infinite swapping limit in Section
\ref{sec:numerical}, highlighting a multiscale integrator proposed in
this work. Numerical tests are presented in Section~\ref{sec:num} to
validate the proposed algorithms.

%%%%%%%%%%%%%%%%%%%%%%%%%%%%%%%%%%%%%%%%%%%%
\section{Theory}\label{sec:theory}

%%%%%%%%%%%%%%%%%%%%%%%%%%%%%%%%%%%%%%%%%%%%

\subsection{Thermal equilibrium average of two-level quantum {systems} and its ring polymer representation}

Consider the thermal equilibrium average of observables as in
\eqref{eq:aveA} for an operator $\wh A$ and a two level Hamiltonian
$\wh H$, where $\beta = (k_B T)^{-1}$ with $k_B$ the Boltzmann
constant and $T$ the absolute temperature. Here, the Hamiltonian is
given by \[ \wh H = \wh T+ \wh V = \frac{1}{2M}
\begin{pmatrix} \wh p^2 & \\ & \wh p^2
\end{pmatrix} +
\begin{pmatrix} V_{00}(\wh q) & V_{01}(\wh q) \\ V_{10}(\wh q) &
V_{11} (\wh q)
\end{pmatrix},
\] where $\wh q$ and $\wh p$ are the nuclear position and momentum
operators, $V(q)$ is a Hermitian matrix for all $q \in \RR^d$, and $M$
is the mass of nuclei (for simplicity we assume all nuclei have the
same mass). The Hilbert space of the system is thus
$L^2(\RR^d) \otimes \CC^2$, where $d$ is the spatial dimension of the
nuclei position degree of freedom, and thus $\tr_{ne}$ in
\eqref{eq:aveA} denotes trace with respect to both the nuclear and
electronic degrees of freedom, namely,
$\tr_{ne}=\tr_{n}\tr_{e}=\tr_{L^2 (\R^d)}\tr_{\C^{2}}$.  The
denominator in \eqref{eq:aveA} is the partition function given by
$\mc{Z} = \tr_{ne}[e^{-\beta \wh H}]$.  For simplicity, we will assume
that the off-diagonal potential functions $V_{01} = V_{10}$ are real
valued and do not change sign.  We also assume that the observable
$\wh A$ only depends on $q$, but it may have off-diagonal elements. As
we shall illustrate in Section~\ref{sec:review}, sampling off-diagonal
elements of $\wh A$ is more challenging with the ring polymer
representation, which will be the focus of our current paper.

{
In this following, we briefly summarize the ring polymer
representation of \eqref{eq:aveA} proposed in our previous work~\cite{LuZhouPIMDSH}, and more details can be found in Appendix \ref{Append:A}.  With the ring polymer representation, the thermal average \eqref{eq:aveA}  can be approximated by  an average with respect to the classical Gibbs
distribution for ring polymers on the extended phase space with
Hamiltonian $H_N$:
\begin{equation}\label{eq:ensembleavgA} \langle \wh{A} \rangle \approx
\frac{1}{(2\pi)^{dN}}\int_{\RR^{2dN}} \ud \bd q \ud \bd p
\sum_{\bd{\ell}\in\{0, 1\}^N} \pi (\widetilde {\bd{z}} )W_N[A]
(\widetilde {\bd{z}}),
\end{equation} with distribution
\begin{equation}\label{eq:pi} \pi (\widetilde {\bd{z}}) =
\frac{1}{\mathcal Z_N} \exp(-\beta_N H_N (\widetilde {\bd{z}})),
\end{equation} 
To simplify the notation, we have denoted by
$\widetilde {\bd z} = (\bd z, \bd\ell) \in \RR^{dN} \times \RR^{dN}
\times \{0, 1\}^N$
a state vector on the extended phase space, where
$\bd z=(\bd q, \bd p)$ are the position and momentum variables.  To be
more specific, $\bd{q} = (q_1, \cdots, q_N)$ and
$\bd{p} = (p_1, \cdots, p_N)$ are the position and momentum of each
bead, and $\bd{\ell} = (\ell_1, \cdots, \ell_N)$ indicates the surface
index of the bead (thus each bead in the ring polymer lives on two
copies of the classical phase space $\RR^{2d}$, see
Figure~\ref{fig:beads} for a schematic plot).  In particular, when
$\ell_k \ne \ell_{k+1}$, two consecutive $k$-th and $(k+1)$-th beads
in the ring polymer stay on different energy surfaces; this will be
referred to as a kink in the ring polymer.  Notice that in
\eqref{eq:pi}, $\mc{Z}_N$ normalizes the distribution in the sense
that
\[ \frac{1}{(2\pi)^{dN}}\int_{\RR^{2dN}} \ud \bd q \ud \bd p
\sum_{\bd{\ell}\in\{0, 1\}^N} \pi (\widetilde {\bd{z}} )=1.
\]
The expressions for $W_N[A]$ and $H_N$ can be found in
Appendix~\ref{Append:A}, and the readers may also refer to
\cite{LuZhouPIMDSH} for detailed derivations.
\begin{figure}[htbp]
\begin{centering}
\includegraphics[scale=0.5]{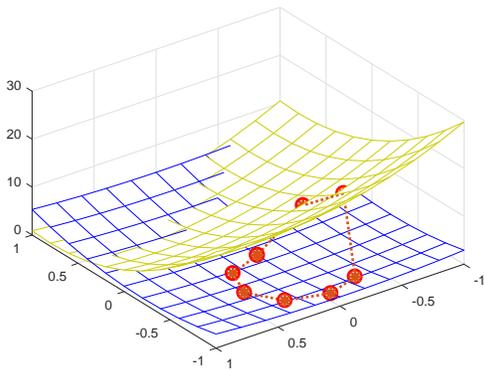}
\caption{Schematic plot of a ring polymer on the extended phase space
with two diabatic energy surfaces.}
\label{fig:beads}
\end{centering}
\end{figure}
}

\subsection{Molecular dynamics sampling of the ring polymer representation}\label{sec:review}

{With the ring polymer representation,} it is then natural to construct a sampling algorithm for $\pi$ as in\eqref{eq:pi} and
thus to approximate the ensemble average in the ring polymer
representation given by \eqref{eq:ensembleavgA}. In our previous work \cite{LuZhouPIMDSH}, a
path-integral molecular dynamics with surface hopping (PIMD-SH) method
was proposed, which we will review here. We also discuss its
difficulty for sampling off-diagonal elements of $\wh{A}$, which
motivates the development of improved algorithms in the current work.

The PIMD-SH method samples the ring polymer representation by
simulating a long trajectory $\widetilde {\bd{z}}(t)$ that is ergodic
with respect to the equilibrium distribution $\pi$, and thus the
ensemble average on the right hand side of \eqref{eq:ensembleavgA} is
approximated by a time average
\begin{equation} \langle \wh{A} \rangle \approx \lim_{T \rightarrow
\infty } \frac{1}{T} \int_0^T W_N [A] (\widetilde {\bd{z}} (t)) \ud t.
\end{equation}

The dynamics of $\wt{\bd{z}}(t)$ is constructed as follows. The
position and momentum part of the trajectory $\bd{z}(t)= (\bd q(t),
\bd p (t))$ evolves according to a Langevin dynamics with Hamiltonian
$H_N (\bd q (t), \bd p(t), \bd\ell (t))$ given the surface index
$\bd\ell (t)$, i.e., a Langevin thermostat is used.  % In other words, the surface 
% , and $\bd z(t)$ follows the Hamiltonian flow on the 
% surface $\bd\ell (t)$ with the Langevin thermostatting.  
More specifically, we have
\begin{equation}\label{eq:mixedham}
  \left\{
 \begin{array}{l l} \ud {\bd q} & = \nabla_{\bd p} H_N (\bd q (t), \bd
p(t), \bd\ell (t)) \ud t, \vspace{0.2cm} \\ \ud {\bd p} & =-
\nabla_{\bd q} H_N (\bd q (t), \bd p(t), \bd\ell (t)) \ud t
\vspace{0.1cm} \\ & \qquad\qquad - \gamma \bd p \ud t + \sqrt{2 \gamma
\beta^{-1}_N M} \ud \bd{B}.
 \end{array} \right.
\end{equation} 
Here $\bd{B}=\bd{B}(t)$ is a vector of $dN$ independent Brownian motions (thus
the derivative of each component is an independent white noise), and
$\gamma \in \mathbb R_+$ denotes the friction constant, as usual in
Langevin dynamics.  By direct calculation, we notice that the evolution of the position just follows as usual
\[ 
\ud {\bd{q}} = \frac{1}{M} \bd{p} \ud t.
\] 
The forcing term of the $\bd{p}$-equation, given by
$-\nabla_{\bd q} H_N (\bd q (t), \bd p(t), \bd\ell (t))$, is in
general $\bd\ell (t)$-dependent, as the potential energy depends on
the level index. The dissipation and fluctuation terms are clearly
independent of $\bd{\ell}(t)$.

The evolution of $\bd{\ell}(t)$ follows a surface hopping type
dynamics in the spirit of the fewest switches surface hopping
\cite{Tully:90, FGASH,FGASH2}, which is a Markov jump process with
infinitesimal transition rate over the time period $(t, t + \delta t)$
for $\delta t \ll 1$ given by
\begin{multline}
  \mathbb{P} \bigl(\bd\ell (t+\delta t) = \bd\ell' \mid \bd\ell (t) = \bd\ell \, , \bd z(t)=\bd z\bigr)
  \\ = \delta_{\bd\ell' , \bd\ell} + \eta \lambda_{\bd\ell', \bd\ell}
  (\bd z) \delta t + o (\delta t).
\end{multline}
Here, $\eta>0$ is an overall scaling parameter for hopping intensity,
and the coefficients $\lambda_{\bd\ell',\bd\ell}$ are given by
\begin{equation} \label{eq:lambda}
  \lambda_{\bd\ell',\bd\ell}(\bd{z}) = 
  \begin{cases}
    \dps
    -  \sum_{\wt{\bd\ell} \in S_{\bd\ell}} p_{\wt{\bd\ell},\bd\ell}(\bd{z}), &  \bd\ell' =\bd\ell,  \\
    p_{\bd\ell',\bd\ell}(\bd{z}), & \bd{\ell}' \in S_{\bd{\ell}},
    \\%|\bd\ell'- \bd\ell|=1\; \text{or} \; \bd\ell'= \bd 1 -\bd\ell, \\
    0, & \text{otherwise}.
\end{cases}
\end{equation}
In the expression above, $S_{\bd\ell}$ denotes the accessible surface
index set from $\bd \ell$ by one ``jump'', which is chosen to be
$S_{\bd\ell}= \{ \bd{\ell}' \mid \norm{\bd\ell'- \bd\ell}_1=1 \;
\mbox{or} \; \bd\ell'= \bd 1 -\bd\ell \}$,
where $\bd 1$ is the vector with all entries $1$. Hence
$\bd{\ell'} = \bd{1} - \bd{\ell}$ indicates that the surface index of
each bead is flipped; and
$\norm{\bd{\ell'} - \bd{\ell}}_1 = \sum_k \abs{\ell'_k - \ell_k} = 1$
indicates that one and only one bead jumps to the opposite energy
surface.  Here in the rate expression, $p_{\bd\ell',\bd\ell} (\bd z)$
is defined as
\begin{equation}\label{eq:rate}
p_{\bd\ell',\bd\ell} (\bd z) = \exp \left( \frac {\beta_N} 2 \bigl(
  H_N(\bd z, \bd\ell) - H_N(\bd z, \bd\ell') \bigr) \right), 
\end{equation}
which is chosen so that the detailed balance relation is satisfied
\begin{multline}\label{eq:detailbalance}
  p_{\bd\ell',\bd\ell} (\bd z ) e^{- \beta_N H_N(\bd z, \bd\ell)}= 
  e^{- \frac {\beta_N} 2 \bigl(
    H_N(\bd z, \bd\ell) + H_N(\bd z, \bd\ell') \bigr)} 
 \\ = p_{\bd\ell,\bd\ell'} (\bd z ) e^{- \beta_N H_N(\bd z, \bd\ell')}.
\end{multline}
This guarantees that the distribution $\pi$ is preserved under the
dynamics of the jumping process. {Moreover, it has been proved in \cite{LuZhouPIMDSH} that  $\pi$ as in \eqref{eq:pi} is indeed the equilibrium distribution of the dynamics $\wt{\bd z}(t)$.}

We conclude this section by pointing out the challenge of the DS
method in sampling off-diagonal elements of $\wh A$.  Observe that,
for $\beta_N \ll 1$, when $\ell_k = \ell_{k+1}$,
$\exp(\beta_N \langle \ell_k| G_k| \ell_{k+1} \rangle)\sim \mathcal
O(1)$,
and when $\ell_k \ne \ell_{k+1}$,
$\exp(\beta_N \langle \ell_k| G_k| \ell_{k+1} \rangle)\sim \mathcal
O(\frac{1}{\beta_N})$.
Thus the asymptotic behavior of the diagonal and off-diagonal matrix
elements of $G_k$ is quite different. As a result, we notice from
\eqref{eq:Gele} that, when $\beta_N \ll 1$ and if we neglect the
energy difference between the surfaces, for fixed $\bd z$,
$H_N(\bd z,\bd \ell)$ is much larger when the ring polymer has more
kinks.  In general, the energy difference cannot be ignored, but still
we observe in implementations, when $\beta_N \ll 1$,
$p_{\bd\ell',\bd\ell} (\bd z)$ might be very small when $\bd \ell'$
has more kinks than $\bd \ell$.

On the other hand, from \eqref{eq:WNA}, we see that only kinks in a
ring polymer effectively contribute to the expectation of off-diagonal
elements of $\wh{A}$. When $\beta_N$ is small, forming kinks in the
trajectory $\wt{\bd{z}}(t)$ is difficult due to the small jumping
rate, as discussed above, but the contribution from each kink to the
expectation is large.  This makes the direct simulation of
$\wt{\bd{z}}(t)$ inefficient to sample off-diagonal observables. We
shall further elaborate this point by comparing the ring polymer
formulation with its infinite swapping limit in Section~\ref{sec:IS}.

One could of course
try to increase the hopping intensity parameter $\eta$ to encourage
hopping of the beads and hence formation of kinks, doing this naively would
unfortunately cause more severe stability constraints on the time
steps in numerically integrating the trajectory
$\wt{\bd{z}}(t)$. Therefore, we aim to explore an alternative
formulation which naturally handles the frequent hopping scenario, and
thus facilitates efficient sampling. This can be achieved
by explicitly taking the infinite swapping limit of $\eta \to \infty$, as
presented in the next section.

\subsection{The infinite swapping limit} \label{sec:IS}

As motivated by the discussions above, we would like to increase the
swapping frequency $\eta$. In fact, as we will discuss further below,
sampling with larger $\eta$ is more efficient. To get around the time
step restriction for the stiffer dynamics with large $\eta$, we
consider in this section the dynamics under the infinite swapping
limit $\eta \to \infty$. The limiting dynamics can actually be
explicitly derived.

When $\eta \gg 1$, the dynamics of $\wt{\bd{z}}(t)$ exhibits a time
scale separation, as the position and momentum degrees of freedom
$\bd z= (\bd q(t), \bd p(t))$ evolve much more slowly than the surface
index $\bd{\ell}(t)$.  In the infinity swapping limit
$\eta \to \infty$, within an infinitesimal time interval, while the
position and momentum $\bd z=(\bd q(t), \bd p(t))$ has barely changed,
the surface index $\bd \ell(t)$ will have already fully explored its
parameter space, and equilibrated to the conditional probability
distribution in $\bd \ell$ with fixed $\bd z$, namely,
\begin{equation}\label{eq:reZz}
\pi (\bd \ell \mid \bd z)= \frac{1}{\mathcal Z_{\bd z}} \exp(-\beta_N H_N (\bd{z}, \bd \ell )).
\end{equation}
Here $\mc{Z}_{\bd z}$ introduced in \eqref{eq:reZz} is the
normalization constant
$\mc{Z}_{\bd z} = \sum_{\bd{\ell} \in \{0, 1\}^N} \exp(-\beta_N H_N
(\bd{z}, \bd{\ell}))$,
such that $\sum_{\bd{\ell}} \pi (\bd \ell \mid \bd z)=1$.  The
equilibration to the conditional distribution is guaranteed by the
detailed balance relation \eqref{eq:detailbalance} for fixed $\bd z$.

As the dynamics of $\bd{\ell}$ is on a much faster scale, the dynamics
of the position and momentum $ (\bd q(t), \bd p(t))$ is driven
effectively by the averaged forcing, given by the average of
$-\nabla_{\bd{q}} H_N(\bd{q}(t), \bd{p}(t), \bd{\ell})$ with respect
to the conditional distribution of $\bd{\ell}$ given
$(\bd{q}(t), \bd{p}(t) )$.  This is the standard averaging phenomenon
for multiscale dynamics and can be made rigorous following for
example the books \cite{PSbook,Ebook}.

To write down the limiting dynamics as $\eta \to \infty$, we calculate the averaged force given $(\bd{q}, \bd{p})$ as
\begin{multline}
  \sum_{\bd{\ell} \in \{0, 1\}^N} \nabla_{\bd{q}} H_N(\bd{q}, \bd{p}, \bd{\ell}) \pi (\bd \ell \mid \bd{q}, \bd{p}) \\
  = \frac{1}{\mc{Z}_{\bd{z}}} \sum_{\bd{\ell} \in \{0, 1\}^N} \nabla_{\bd{q}} H_N(\bd{q}, \bd{p}, \bd{\ell}) e^{-\beta_N H_N(\bd{q}, \bd{p}, \bd{\ell})}.
\end{multline}
To simplify the notation, let us define an averaged Hamiltonian as
\begin{equation}
  \wb{H}_N(\bd{q},\bd{p})= -\frac{1}{\beta_N} \ln \Bigl(\sum_{\bd{\ell} \in \{0, 1\}^N}  e^{-\beta_N H_N(\bd{q}, \bd{p}, \bd{\ell})} \Bigr), 
\end{equation}
such that 
\begin{equation}
  \mc{Z}_{\bd{z}} = \exp(-\beta_N \wb{H}_N(\bd{q}, \bd{p}) ).
\end{equation}
We observe by direct calculation that 
\begin{equation}
  \nabla_{\bd{q}} \wb{H}_N(\bd{q}, \bd{p}) = \frac{1}{\mc{Z}_{\bd{z}}} \sum_{\bd{\ell} \in \{0, 1\}^N} \nabla_{\bd{q}} H_N(\bd{q}, \bd{p}, \bd{\ell}) e^{-\beta_N H_N(\bd{q}, \bd{p}, \bd{\ell})}, 
\end{equation}
and hence the average forcing with respect to the conditional distribution $\pi(\bd{\ell} \mid \bd{q}, \bd{p})$ is given by 
\begin{equation}
  - \nabla_{\bd{q}} \wb{H}_N(\bd{q}, \bd{p}) = - \sum_{\bd{\ell} \in \{0, 1\}^N} \nabla_{\bd{q}} H_N(\bd{q}, \bd{p}, \bd{\ell}) \pi(\bd \ell \mid \bd{q}, \bd{p}).
\end{equation}
As a result, in the infinite swapping limit $\eta \to \infty$, the Langevin dynamics  \eqref{eq:mixedham} takes the limit 
\begin{align} \label{eq:qa}
  {\ud \bd q}&=  \frac{\bd p}{M} \ud t. \\
  \label{eq:pa}
  {\ud \bd p}&= - \nabla_{\bd q} \wb{H}_N(\bd q, \bd p) \ud t - \gamma \bd p \ud t + \sqrt{2 \gamma \beta_N^{-1}M} \ud \bd B.
\end{align}
Here as in \eqref{eq:mixedham} $\bd B = \bd B(t)$ is a vector of $dN$
independent Brownian motion and $\gamma \in \mathbb R^+$ denotes the
friction constant.

{The limiting dynamics \eqref{eq:qa}-\eqref{eq:pa} samples the invariant measure 
\begin{equation}\label{eq:traceprod}
  \begin{aligned}
  \wb{\pi}(\bd{q}, \bd{p}) & \propto \exp(-\beta_N \wb{H}_N(\bd{q}, \bd{p}))\\
  & = \exp\Bigl(-\beta_N \tr \bigl( \prod_{k=1}^N G_k \bigr) \Bigr),
\end{aligned}
\end{equation}
where the expression of $G_K$ is given in the
Appendix~\ref{Append:A}. Note that this coincides with the equilibrium
measure of the mean-field ring polymer configuration
\cite{PrezhdoRossky:97,DukeAnanth2016}. Thus the infinite swapping
limit of the path integral molecular dynamics with surface hopping can
be understood as the mean-field path integral molecular dynamics.}

The average of observable in the ring polymer representation can be
also rewritten using the conditional distribution
$\pi(\bd\ell \mid \bd z)$.  Recall the expectation formula for an
observable $\wh{A}$ in the ring polymer representation 
\eqref{eq:WNA}.
\begin{align*}
  \tr_{ne}\,& [e^{-\beta \wh H}\wh A ] \\
  & \approx  \int_{\RR^{2dN}} \frac{\ud \bd q \ud \bd p}{(2\pi)^{dN}}\,  
    \sum_{\bd{\ell} \in \{0, 1\}^N} \exp(-\beta_N H_N) W_N[A]  \\
  & = \int_{\RR^{2dN}} \frac{\ud \bd q \ud \bd p}{(2\pi)^{dN}}\, 
    \mc{Z}_{\bd{z}} \sum_{\bd{\ell} \in \{0, 1\}^N} \frac{\exp(-\beta_N H_N)}{\mc{Z}_{\bd{z}}} W_N[A] \\
  & = \int_{\RR^{2dN}} \frac{\ud \bd q \ud \bd p}{(2\pi)^{dN}}\, e^{-\beta_N \wb{H}_N(\bd{z})} 
    \sum_{\bd{\ell} \in \{0, 1\}^N} \pi(\bd\ell \mid \bd z) W_N[A].
\end{align*}
Thus, if we define the weighted averaged observable 
\begin{equation} \label{eq:aWA}
\widetilde W [A](\bd z) := \sum_{\bd{\ell} \in \{0, 1\}^N} \pi(\bd\ell \mid \bd z) W_N[A](\bd z, \bd \ell), 
\end{equation}
the expectation is approximated by 
\begin{equation}
  \tr_{ne}[e^{-\beta \wh H}\wh A ]  = \frac{1}{(2\pi)^{dN}} \int_{\RR^{2dN}} \ud \bd q \ud \bd p \, e^{-\beta_N \wb{H}_N(\bd{z})} \widetilde W [A](\bd z).
\end{equation}
In the infinite swapping limit, as the dynamics of $\bd\ell$ has
instantaneously reached to the equilibrium given
$\bd z = (\bd q, \bd p)$, the configuration space of the limiting
dynamics only consists of the position and momentum $(\bd q,\bd
p)$.
Therefore, in the corresponding sampling via the infinite swapping
dynamics, it suffices to evolve $(\bd q, \bd p)$ equations
\eqref{eq:qa}--\eqref{eq:pa} and approximate the ensemble average by
the time average as
\[
\langle \wh A \rangle \approx  \lim_{T \rightarrow \infty} \frac{1}{T} \int_0^T \widetilde W [A] (\bd z (t)) \ud t.
\]
This is the foundation of the numerical sampling algorithm based on the infinite swapping limit. Base on this, we propose efficient and stable algorithms for sampling thermal averages in Section \ref{sec:numerical}.

\medskip 

Before we conclude this section and turn to numerical sampling
schemes, let us give some theoretical analysis of the efficiency of
the infinite swapping limit.  By \eqref{eq:aWA}, we have
\begin{equation}\label{eq:exp} \bigl\langle W_N[A](\bd z, \bd \ell)
  \bigr\rangle = \bigl\langle \wt W_N[A]( \bd z) \bigr\rangle,
\end{equation}
where the first average is taken over the configurational space of
$(\bd z, \bd\ell)$, while the second one is taken over the space of
$\bd z$. Let us now compare the variance of the two estimators, and
calculate
\begin{multline}\label{eq:esqr1} 
  \bigl\langle (W_N[A](\wt {\bd z }))^2\bigr \rangle =
  \int_{\RR^{2dN}} \frac{\ud \bd z}{(2\pi)^{dN}}
  \sum_{\bd{\ell}\in\{0,
    1\}^N} \pi (\widetilde {\bd{z}} ) (W_N[A](\wt {\bd z }) )^2 \\
  = \int_{\RR^{2dN}} \frac{\ud \bd z}{(2\pi)^{dN}} \pi(\bd{z})
  \sum_{\bd\ell} \pi(\bd\ell \mid \bd z) (W_N[A](\wt {\bd z }) )^2,
\end{multline}
where $\pi(\bd{z})$ is the marginal distribution of $\pi(\bd z, \bd \ell)$ on $\bd z$. On the other hand, we have for the estimator $\wt{W}_N[A]$ 
\begin{multline}\label{eq:esqr2} 
  \bigl\langle \wt W_N[A]({\bd z }) )^2\bigr\rangle = \int_{\RR^{2dN}}
  \frac{\ud \bd z}{(2\pi)^{dN}}
  \pi(\bd z) ( \wt W_N[A]({\bd z }) )^2  \\
  = \int_{\RR^{2dN}} \frac{\ud \bd z}{(2\pi)^{dN}} \pi(\bd z) \Bigl(
  \sum_{\bd \ell} W_N[A](\wt {\bd z }) \pi(\bd \ell \mid \bd z)
  \Bigr)^2.
\end{multline}
Therefore, by Jensen's inequality (recall that
$\pi(\bd\ell \mid \bd z)$ is a conditional probability on $\bd\ell$):
\begin{equation}
  \sum_{\bd\ell} \bigl(W_N[A](\wt {\bd z }) \bigr)^2 
  \pi(\bd\ell \mid \bd z) \geq  
  \Bigl( \sum_{\bd\ell} W_N[A](\wt {\bd z })  \pi(\bd\ell \mid \bd z) 
  \Bigr)^2,
\end{equation}
this implies that 
\begin{equation}
  \bigl\langle (W_N[A](\wt {\bd z }))^2\bigr\rangle  \geq \bigl\langle ( \wt W_N[A]({\bd z }) )^2\bigr\rangle.
\end{equation}
Thus, we arrive at 
\begin{equation}
  \Var \bigl(W_N[A](\wt{ \bd z}) \bigr)  \geq \Var \bigl(\wt W_N[A](\bd z) \bigr).
\end{equation}
This means that the estimator of the infinite swapping limit has a
smaller variance than that of the original ring polymer
representation.  Moreover, in terms of the convergence to invariant
measure, using the large deviation theory, it can be proved that the
dynamics with a larger $\eta$ has faster convergence, and hence the
infinite swapping limit is also superior. The rigorous analysis
follows similar argument as in \cite{Dupuis2012,
  LuVEremd}. Intuitively, the faster convergence is easy to understand
since a larger $\eta$ accelerates the sampling in the $\bd\ell$
variable and in the infinite swapping limit, the average over
$\bd\ell$ is explicitly taken.  The smaller variance and fast
convergence to equilibrium justify the use of the infinite swapping
limit.

\section{Numerical methods} \label{sec:numerical}

\subsection{Simulation of the PIMD-SH dynamics and its infinite swapping limit}

For completeness, let us first summarize the steps of direct
simulation of the PIMD-SH method \cite{LuZhouPIMDSH}. For the initial
conditions to the trajectory
$\widetilde {\bd{z}} (0)=(\bd q (0), \bd p(0), \bd\ell(0))$, thanks to
the ergodicity of the dynamics, any initial conditions can be in
principle used, while a better initial sampling will accelerate the
convergence of the sampling. In our current implementation, for
simplicity, we initialize all the beads in the same position, sample
their momentum according to a Gaussian distribution
$\mathcal N (0, M \beta_N^{-1})$, and take $\bd\ell(0) = \bd{0}$,
where $\bd{0}$ is a vector of all zeros, meaning that initially all
beads of the ring polymer stay on the lower energy surface.

The overall strategy we take for the time integration is time
splitting schemes, by carrying out the jumping step, denoted by
$\textsf{J}$, and the Langevin step denoted by $\textsf{L}$, in an
alternating way.  In this work, we apply the Strang splitting, such
that the resulting splitting scheme is represented by
$\textsf{JLJ}$. This means that, within the time interval
$[ t^n, t^n +\Delta t]$ ($\Delta t$ being the time step size), we
carry out the following steps in order:
\begin{enumerate}
\item Numerically simulate the jumping process for $\bd{\ell}$ for
  $\Delta t /2$ time with fixed position and momentum of the ring polymer;

\item Propagate numerically the position and momentum of the ring
  polymer using a discretization of the Langevin dynamics for
  $\Delta t$ time while fixing the surface index $\bd\ell$ (from the previous sub-step);

\item The jumping process for $\bd{\ell}$ is simulated for another $\Delta t/2$ time
  with fixed position and momentum of the ring polymer;

\item The weight function $W_N[A](\widetilde {\bd{z}} (t^{n+1}))$ of
  the observable $\wh{A}$ is calculated (and stored, if needed) to update the
  running average of the observable.
\end{enumerate}
The above procedure is repeated for each time step until we reach a
prescribed total sampling time $T$ or when the convergence of the
sampling is achieved under certain stopping criteria {(for
  example, when the estimated empirical variance is smaller than a
  prescribed threshold)}.  In our test examples, we use standard
stochastic simulation algorithm (kinetic Monte Carlo scheme)
\cite{Gillespie} for the jumping process and the BAOAB integrator for
the Langevin dynamics \cite{LeimkuhlerMatthews}, the details of both
can be found in \cite{LuZhouPIMDSH}. As we already mentioned above,
when the swapping frequency $\eta$ is large, we need to take very
small time step in the splitting scheme to ensure stability.

The extension of the numerical methods to the infinite swapping limit
is natural, which we shall refer to as the straightforward simulation
method of the infinite swapping limit (abbreviated by IS
hereinafter). In the infinite swapping limit, since we no longer keep
track of the discrete level variable $\bd \ell$ (whose effect is
averaged out), in the initialization step, we only need to specify
$\bd z (0)= \big( \bd q(0),\bd p(0) \bigr)$. Again, any choice of the
initial condition can be taken due to ergodicity.

For the numerical integration, we again use the time splitting
scheme. In the infinite swapping step, denoted by
$\textsf{J}_{\text{inf}}$, we compute the conditional distribution
$\pi (\bd \ell \mid \bd z)$ as in \eqref{eq:reZz} for all possible
$\bd \ell$ with fixed $(\bd q, \bd p)$. For the averaged Langevin
step, denoted by $\textsf{L}_{\text{inf}}$ we evolve the averaged
Langevin dynamics \eqref{eq:qa}, \eqref{eq:pa} with fixed conditional
distribution $\pi (\bd \ell \mid \bd z)$. In this work, we choose to
use the symmetric Strang splitting represented by
$\textsf{J}_{\text{inf}}\textsf{L}_{\text{inf}}\textsf{J}_{\text{inf}}$.
In each time interval $[ t^n, t^n +\Delta t]$ ($\Delta t$ being the
time step size), we carry out the following steps in order:
\begin{enumerate}
\item We compute the conditional distribution as in \eqref{eq:reZz}
  with fixed position and momentum of the ring polymer. Note that
  except for the first iteration, this step can be skipped since the
  conditional distribution has already been obtained in the previous
  time step;

\item We propagate numerically the position and momentum of the ring
  polymer using the BAOAB discretization of the averaged Langevin
  dynamics \eqref{eq:qa}, \eqref{eq:pa} for $\Delta t$ time with the
  fixed conditional distribution (from the previous sub-step);

\item The averaged weight function
  $\widetilde W_N[A](\bd{z} (t^{n+1}))$ of the observable $\wh{A}$ is
  calculated using the conditional distribution {as in \eqref{eq:reZz}} from the previous
  sub-step to update the running average of the observable.
\end{enumerate}
The above procedure is repeated for each time step until we reach a
prescribed total sampling time $T$ or when the convergence of the
sampling is achieved under certain stopping criteria. We remark that,
the symmetric splitting is almost as cheap as a first order splitting,
since the first $\textsf{J}_{\text{inf}}$ sub-step in each evolution
loop
$\textsf{J}_{\text{inf}}\textsf{L}_{\text{inf}}\textsf{J}_{\text{inf}}$
can be skipped from the second iteration.

{As we will show in Section~\ref{sec:num} that the infinite
  swapping PIMD-SH (or equivalently the mean-field PIMD) is better
  than the direct simulation of PIMD-SH. When $N$ is very large, while
  the calculation of the averaged force and weighted averaged
  observable is possible by utilizing the trace product formula as in
  \eqref{eq:traceprod}, it could be numerically unstable as it
  involves multiplications of large number of matrices. In the next
  section, we show that even though all possible level configurations
  grows exponentially as $2^N$, if we view that as a sampling problem in the
  configuration space of $\bd{\ell}$, it is possible to devise
  efficient sampling schemes to circumvent enumerating all possible
  configurations. This provides an alternative efficient method to simulate the infinite swapping limit.}

\subsection{A multiscale implementation of the infinite swapping
  limit}\label{sec:hmm}

As we have pointed out above, the direct simulation of PIMD-SH becomes
expensive when $\eta \to \infty$ due to the time step size
restrictions. The difficulty can be understood as due to the huge
time-scale separation of the dynamics of $\bd{\ell}$ and
$(\bd{q}, \bd{p})$ when $\eta \gg1$, as the fast time scale of
$\bd{\ell}$ restricts the time step size. To deal with such scale
separation, in this section, {we propose a multiscale integrator
  for efficient simulation of the infinite swapping limit following
  the spirit of heterogeneous multiscale method (abbreviated by HMM
  hereinafter) \cite{EEngquist:03,EVE:03,Weinan2007,Ebook}. Similar
  idea of exploiting multiscale integrator has been also used by one
  of us for replica exchange method in \cite{TangLuAbramsVE, LuVEremd}
  and for irreversible Langevin sampler in \cite{LuSpiliopoulos}. Let
  us also mention that the multiscale integrator ideas have been
  proposed in various fields, e.g., early works in the context of
  linear multistep methods for ODEs \cite{GearWells84} and for
  molecular dynamics with multiple time scales \cite{Tuckerman1990,
    TuckermanBerne1991, TuckermanBerne1991_2,TuckermanBerne1991_3}}.

For the HMM scheme, one evolves the slow dynamics of  $(\bd{q}, \bd{p})$ 
using a macrosolver, while the fast dynamics of $\bd \ell$ (as $\eta \gg 1$) 
is evolved using a microsolver. The necessary input of the macrosolver (in this case, the averaged force) is obtained from the microsolver through an estimator. The HMM schemes consists of the microsolver, the macrosolver and the estimator connecting the two. 

More specifically, here we will use a BAOAB splitting scheme as the
macrosolver to evolve the averaged Langevin dynamics of
$(\bd{q}, \bd{p})$, where the weighted sum in the force term is
replaced by the approximation provided by the estimator. Choose an
appropriate macro time step $\Delta t$ and a frequency $\eta$ such
that $\eta \gg \frac{1}{\Delta t}$, for example
$\eta = \frac{R}{\Delta t}$, where $R =10^1 \sim 10^3$ is understood
as the ratio between the number of macrosteps for $\bd{z}$ and
microsteps for $\bd\ell$. The overall algorithm goes as following for
each macro time step $k$:
\begin{enumerate}
\item Microsolver. Evolve $\bd \ell_k$ via a stochastic simulation
  algorithm from $t_k$ to $t_{k+1}:=t_k + \Delta t$ using the rate in
  \eqref{eq:lambda}. That is, set $\bd \ell_{k,0}=\bd \ell_k$,
  $t_{k,0}=t_k$, and for $j \ge 1$, do
  \begin{enumerate}
  \item Compute the lag via
    \[ \tau_j = - \frac{\ln r}{\eta \displaystyle \sum_{\wt{\bd\ell}
        \in S_{\bd\ell_{k,j-1}}}
      p_{\wt{\bd\ell},\bd\ell_k,j-1}(\bd{z_k})},
    \]
    where $r$ is a random number uniformly distributed in the interval $(0,1)$.
  \item Pick $\bd \ell_{k,j} \in S_{\bd\ell_{k,j-1}} $ with
    probability
    \[
    \frac{ p_{{\bd\ell_{k,j}},\bd\ell_k,j-1}(\bd{z_k})}{ \displaystyle \sum_{\wt{\bd\ell} \in S_{\bd\ell_{k,j-1}}}  p_{\wt{\bd\ell},\bd\ell_k,j-1}(\bd{z_k})}.
    \]
  \item Set $t_{k,j}=t_{k,j-1}+\tau_j$ and repeat till the first $J$,
    such that $t_{k,J}>t_k+\Delta t$. Then, set
    $\bd \ell_{k+1}=\bd \ell_{k,J}$ and reset
    $\tau_J=t_k + \Delta t - t_{k,J-1}$.
  \end{enumerate}

\item Estimator. Given the trajectory of $\bd \ell$, namely,
  $\bd \ell_{k,1},\cdots \bd \ell_{k,J}$ associated with
  $\tau_1,\cdots,\tau_J$. We estimate the averaged force term by
  \begin{multline} \label{eq:aforce} -\nabla_{\bd q}
    \wb{H}(\bd{z})=- \sum_{\bd{\ell} \in \{0, 1\}^N} \pi (\bd
    \ell \mid \bd z) \nabla_{\bd q} H_N \\ \approx - \frac{1}{\Delta
      t} \sum_{j=1}^J \nabla_{\bd q} H_N ( \bd z, \bd \ell_{k,j})
    \tau_j.
  \end{multline}
  And the weighted average is approximated by
  \begin{equation}
    \widetilde W [A] (\bd z_k) \approx  \frac{1}{\Delta t}  \sum_{j=1}^J  W_N[A](\bd z_k,\bd \ell_{k,j}) \tau_j.
  \end{equation}

\item Macrosolver. Evolve $\bd z_k$ to $\bd z_{k+1}$ using one
  time-step of size $\Delta t$ using BAOAB integrator for the Langevin
  equations with the force term replaced by \eqref{eq:aforce}
  calculated in the estimator.
\end{enumerate}
The above three steps are repeated till the final simulation time.

When $R$ is sufficiently large, the microsolver and the estimator will
give an accurate estimation of the averaged force since the jumping
process is ergodic with respect to the conditional probability due to
detailed balance. Hence the HMM integrator effectively simulates the
infinite swapping limit. The numerical analysis of the scheme follows
standard machinery of HMM type integrators for multiscale dynamics,
see e.g., \cite{EEngquist:03, EVE:03, ELiuVE:05, TaoOwhadiMarsden2010,
  LuSpiliopoulos}, and we will not go into the details here. Rather,
we shall investigate numerically in Section~\ref{sec:num} the
efficiency of the integrator and choice of parameters.

\section{Numerical Tests} \label{sec:num}

\begin{figure}[htbp]
\begin{center}
\includegraphics[scale=0.55]{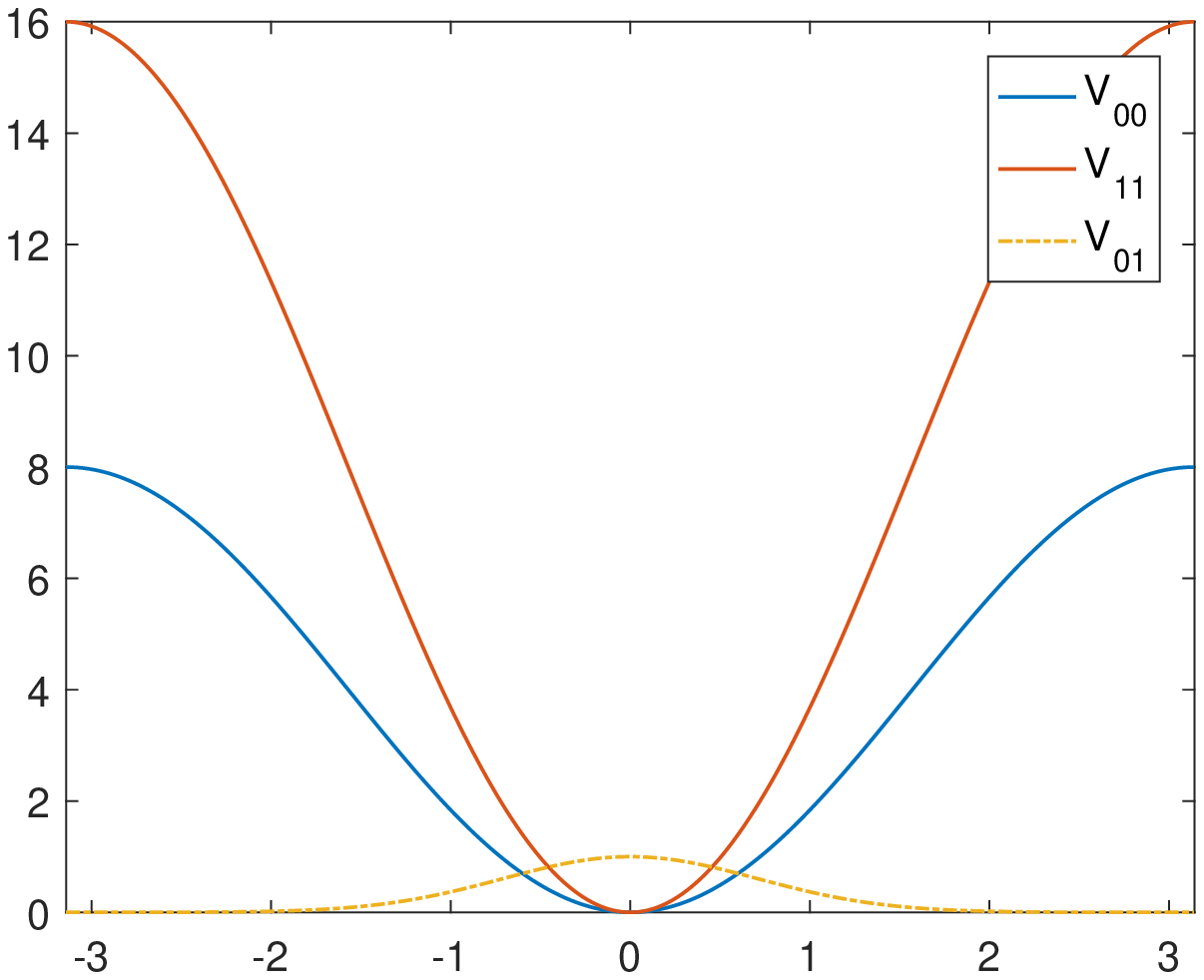}
\includegraphics[scale=0.55]{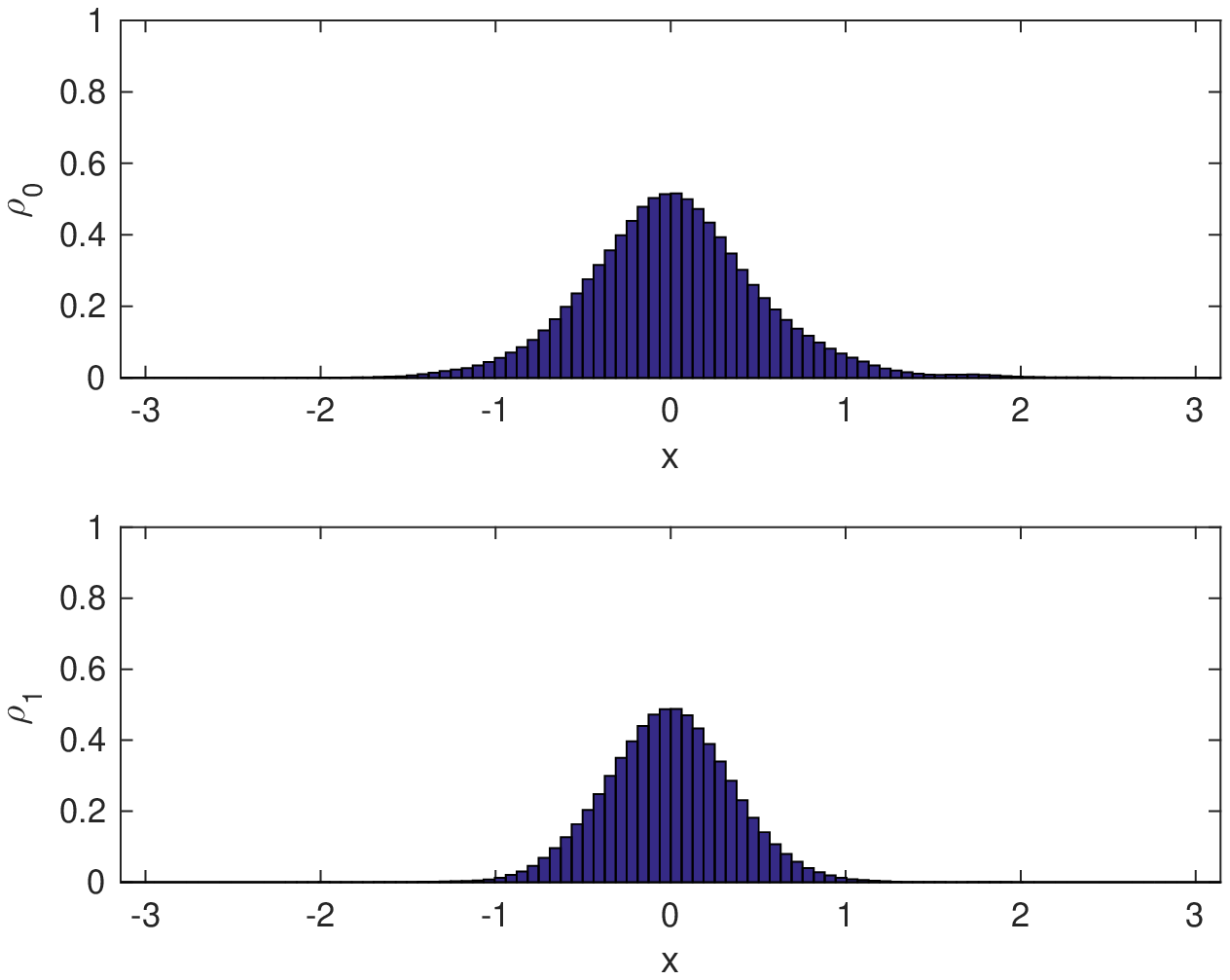} \\
\caption{ Top: diabatic potential surfaces for the test example
  \eqref{ex:pot1}. Bottom: equilibrium distribution on both
  surfaces.  }
\label{fig:Eplot1}
\end{center}
\end{figure}

To validate the PIMD-SH method in the infinite swapping limit, we
consider test problems with the following two potentials. Both
potentials are chosen to be one-dimensional and periodic over
$[-\pi,\pi]$, so that the reference solutions can be obtained
accurately with pseudo-spectral approximations and compared to PIMD-SH
results. The first test potential is given
by %\jl{do we ever choose $\delta$ that is not $1$?}
\begin{equation}  \label{ex:pot1}
\left \{
\begin{split}
V_{00} &=a \bigl(1-\cos(x) \bigr); \\
V_{11} & = b\bigl(1-\cos(x) \bigr);\\
V_{01} & = V_{10} =  c e^{-d x^2}.
\end{split}
\right.
\end{equation}
We take $b>a$, so $V_{11} \ge V_{00}$ and the two energy surfaces only
intersect at $x=0$, where the off-diagonal term takes its largest
value.  %The parameter $\delta$ determines the coupling strength.
The energy surfaces are symmetric with respect to $x = 0$. At thermal
equilibrium, the density is expected to concentrate around $x=0$,
where transition between the two surfaces is the most noticeable due
to the larger off-diagonal coupling terms. In this work, we
choose $a=4$, $b=8$, $c=1$ and $d=1$.  The diabatic energy surfaces
with equilibrium distributions on each surface are plotted in Figure~\ref{fig:Eplot1}.

\begin{figure}[htbp]
\begin{center}
\includegraphics[scale=0.55]{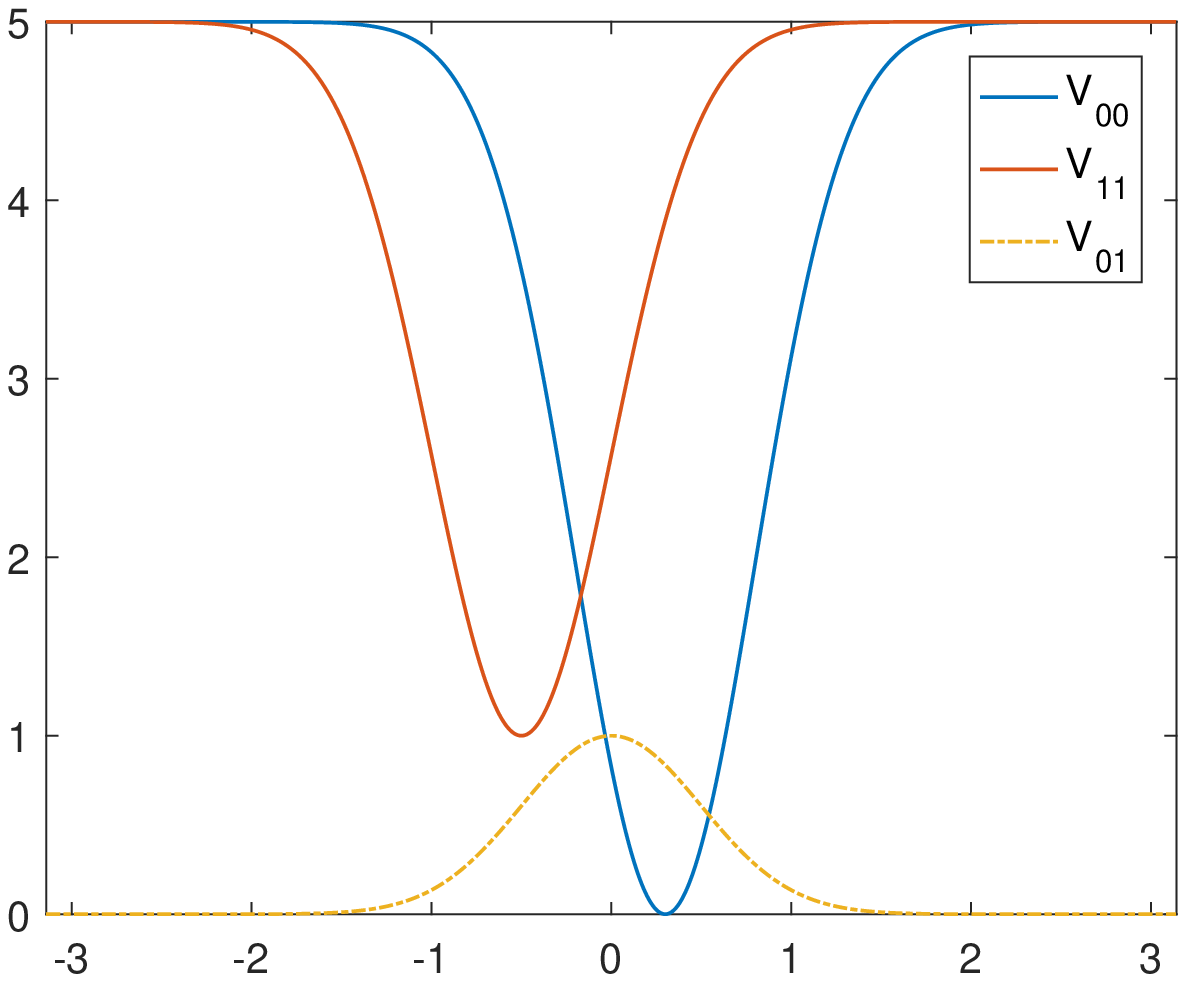} 
\includegraphics[scale=0.55]{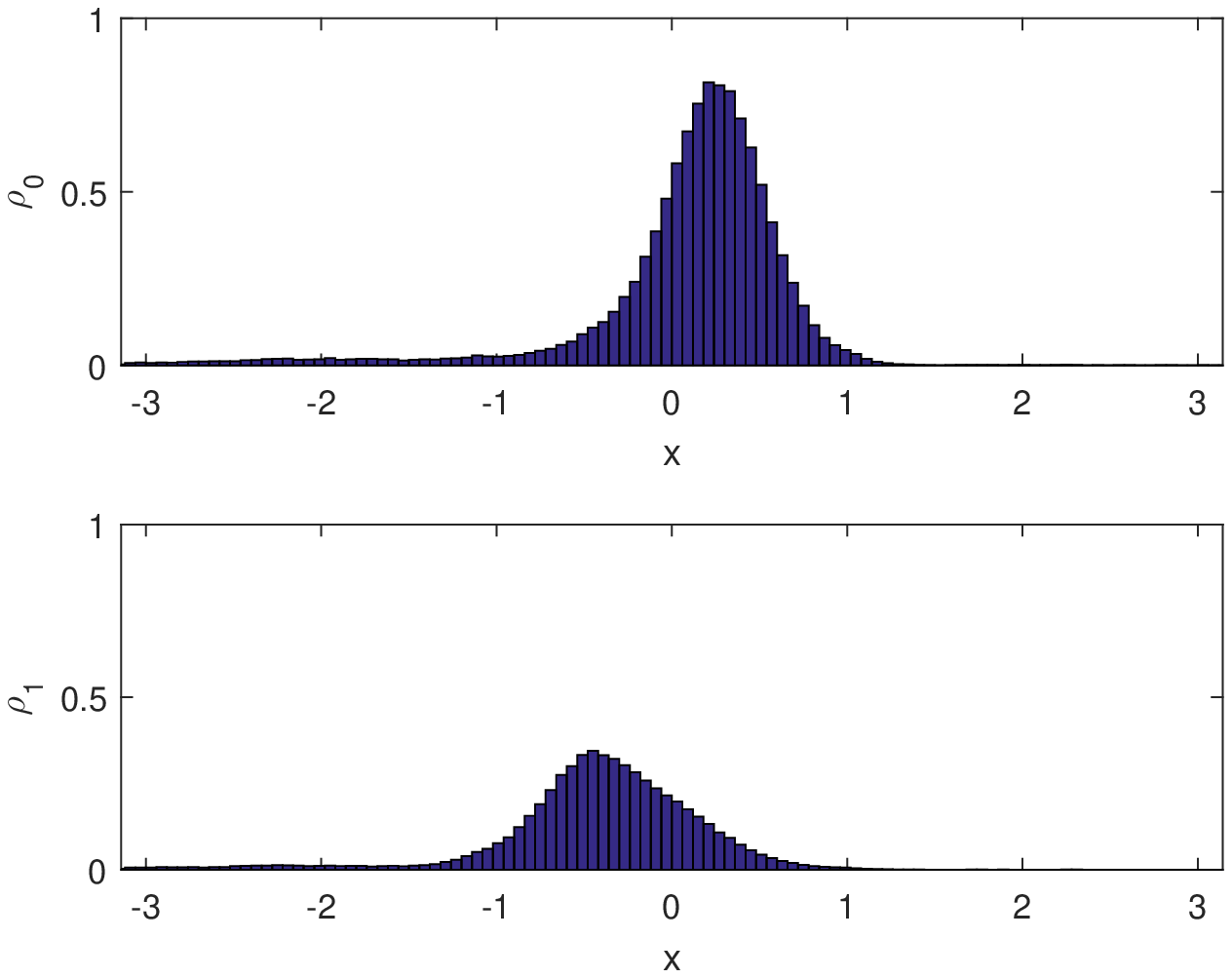} \\
\caption{Left: diabatic potential surfaces for the test example
  \eqref{ex:pot2}. Right: equilibrium distribution on both
  surfaces. }
\label{fig:Eplot2}
\end{center}
\end{figure}

The other test potential we take is given by 
\begin{equation}  \label{ex:pot2}
\left \{
\begin{split}
V_{00} &= a_1-b_1 e^{-2(x-c_1)^2}; \\
V_{11} & =a_2-b_2 e^{-2(x-c_2)^2};\\
V_{01} & = V_{10} = d_1 e^{-d_2 x^2}.
\end{split}
\right.
\end{equation}
$V_{00}$ and $V_{11}$ take their minima at $x=c_1$ and $x=c_2$
respectively, while the off-diagonal is peaked around $x=0$. In this
model, the potential is asymmetric, and the two potentials have
different minima, which also deviate from where the surface hopping is
the most active due to the off-diagonal potential. These make this
test model more challenging than the previous one. In this paper, we
choose $a_1=5$, $a_2=5$, $b_1=5$, $b_2=4$, $c_1=0.3$, $c_2=-0.5$,
$d_1=1$ and $d_2=2$. We plot the diabatic energy surfaces and the
equilibrium distributions on each surface in Figure~\ref{fig:Eplot2}.

For both potential examples, we test and compare the performances of
numerical methods with the diagonal observable
\begin{equation} \label{eq:ob1}
\wh A=\begin{bmatrix}
  e^{- {\wh q}^2}   &  0\\
 0 &   e^{- {\wh q}^2}  
\end{bmatrix},
\end{equation}
and also the off-diagonal observable
\begin{equation} \label{eq:ob2}
\wh{A}=\begin{bmatrix}
0   &    e^{- {\wh q}^2}\\
  e^{- {\wh q}^2} &   0  
\end{bmatrix}.
\end{equation}

\subsection{Tests for diagonal and off-diagonal observables with
  different $\Delta t$}

Let us first use potential model \eqref{ex:pot1} in order to compare
with our previous method. We test the IS method and the HMM method for
the number of beads $N=4$, $\beta=1$ and $\Delta t = \frac{1}{20}$,
$\frac{1}{40}$, $\frac{1}{80}$ and $\frac{1}{160}$, which are compared
with the direct simulation (DS) method.  The tests are implemented
with both the diagonal observable \eqref{eq:ob1} and the off-diagonal
observable \eqref{eq:ob2}.

\begin{figure}[htbp]
\begin{center}
\includegraphics[scale=0.55]{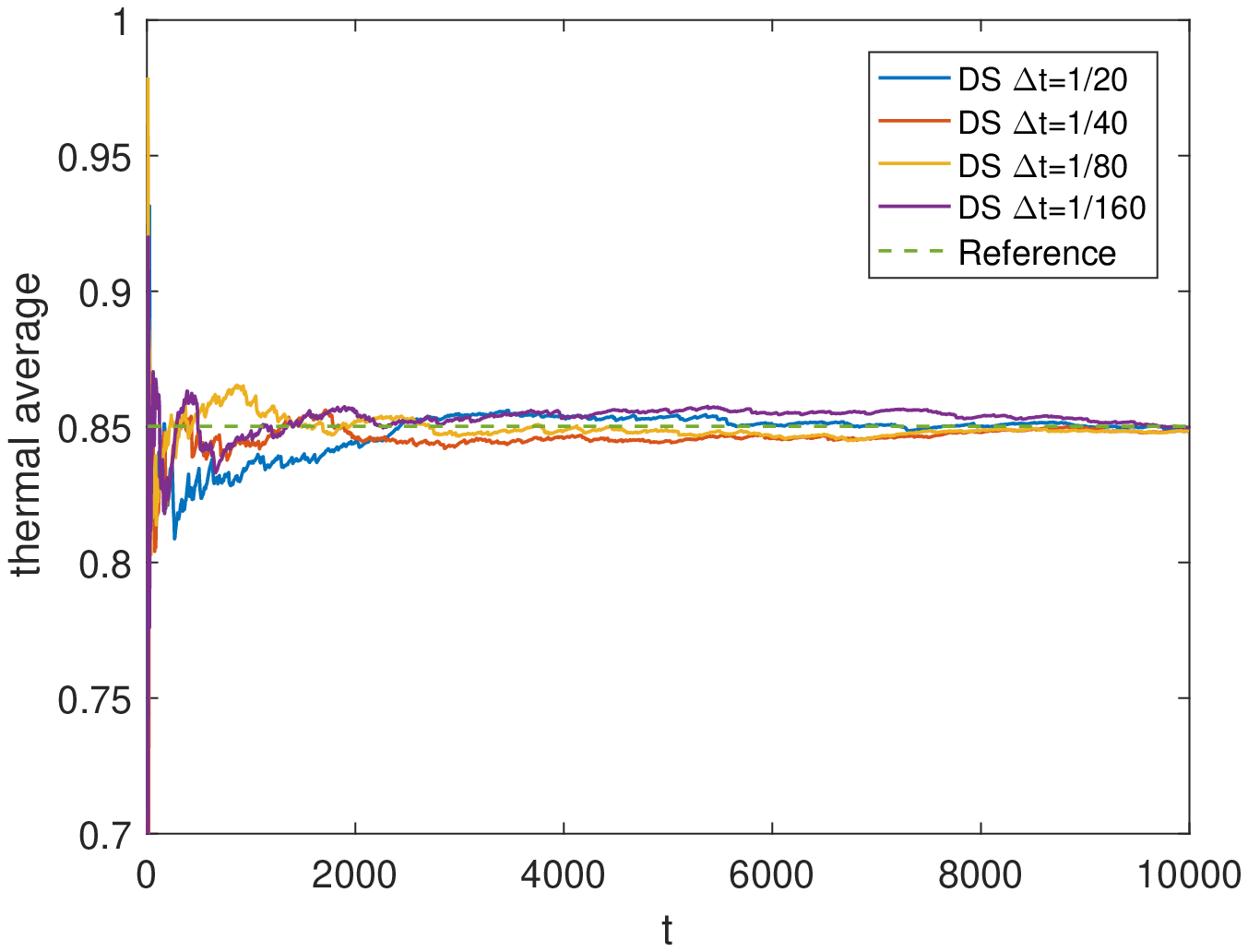} \includegraphics[scale=0.55]{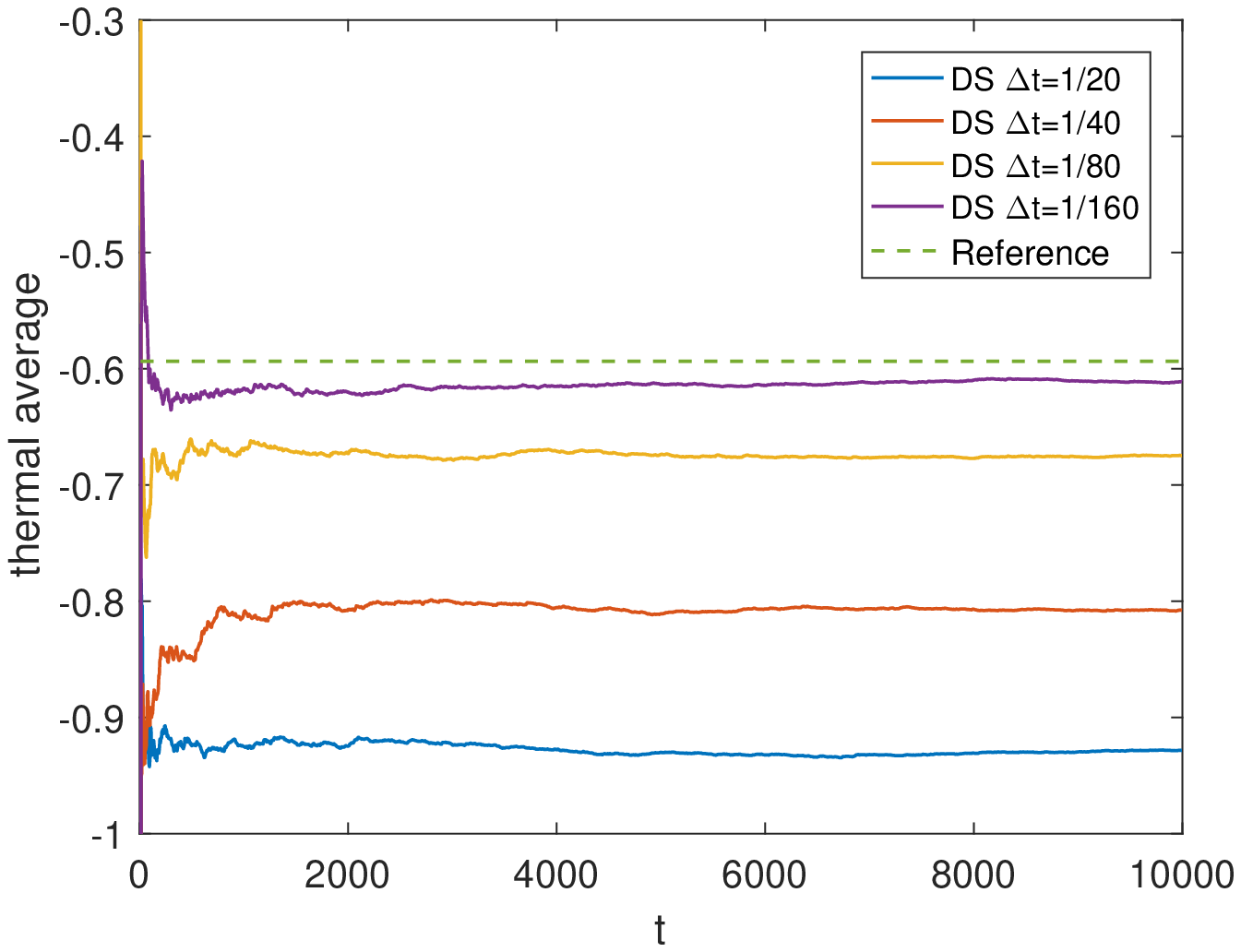}\\
\caption{ DS method with $N=4$, $\beta=1$ and various $\Delta t$ with
  potential model \eqref{ex:pot1}. Top: diagonal observable, the
  reference value is $0.850241$. Bottom: off-diagonal observable, the
  reference value is $-0.593497$.}
\label{fig:PIMDSHdt}
\end{center}
\end{figure}

In Figure~\ref{fig:PIMDSHdt}, we plot the numerical results with the
DS method. We observe that for the diagonal observable \eqref{eq:ob1},
we cannot clearly distinguish the performance with different time step
sizes $\Delta t$ and all the tests are able to capture the thermal
averages accurately, but for the off-diagonal observable
\eqref{eq:ob2}, we need to take sufficiently small $\Delta t$ in order
to obtain a reliable simulation. This confirms our observations in the
previous paper.

\begin{figure}[htbp]
\begin{center}
\includegraphics[scale=0.55]{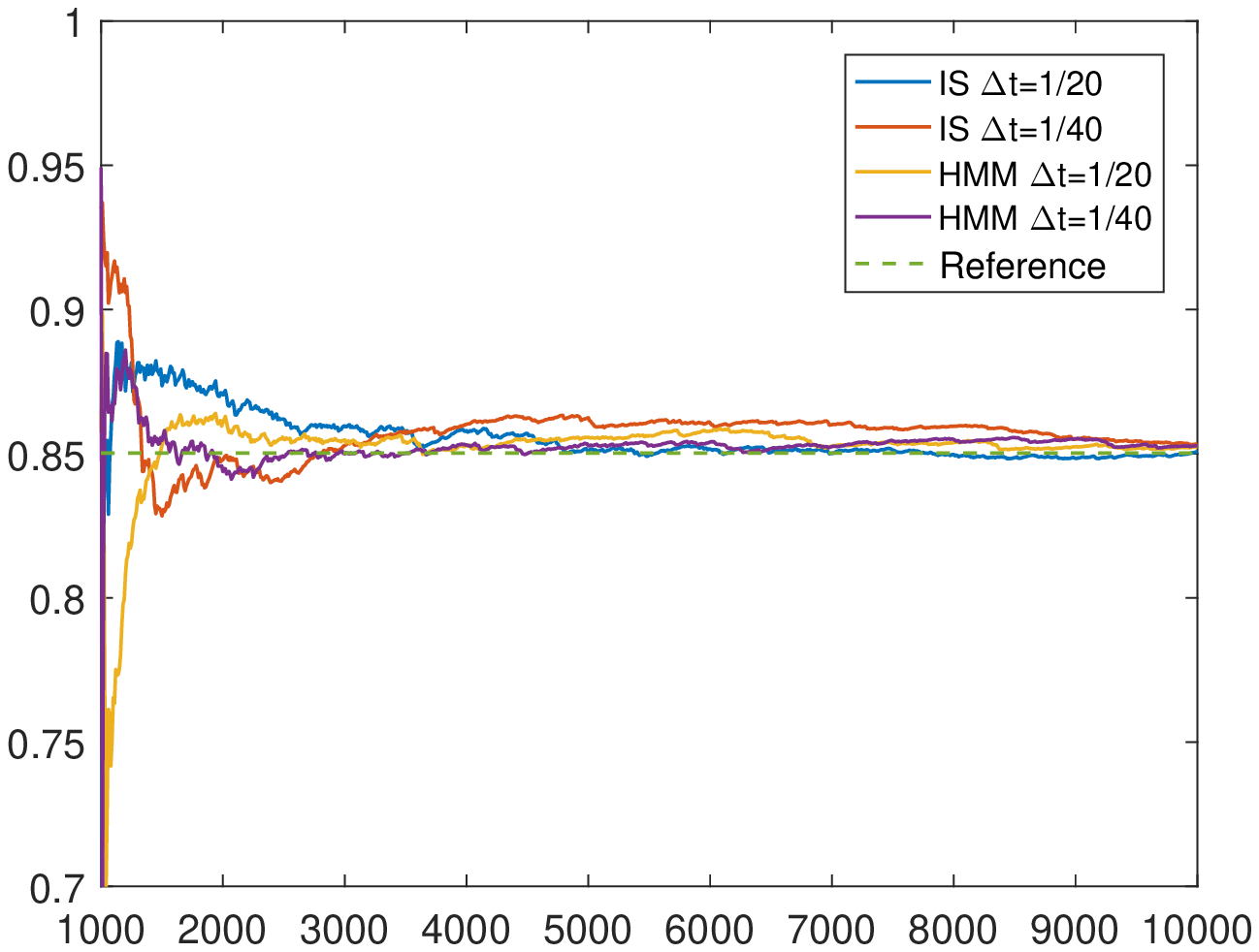} \includegraphics[scale=0.55]{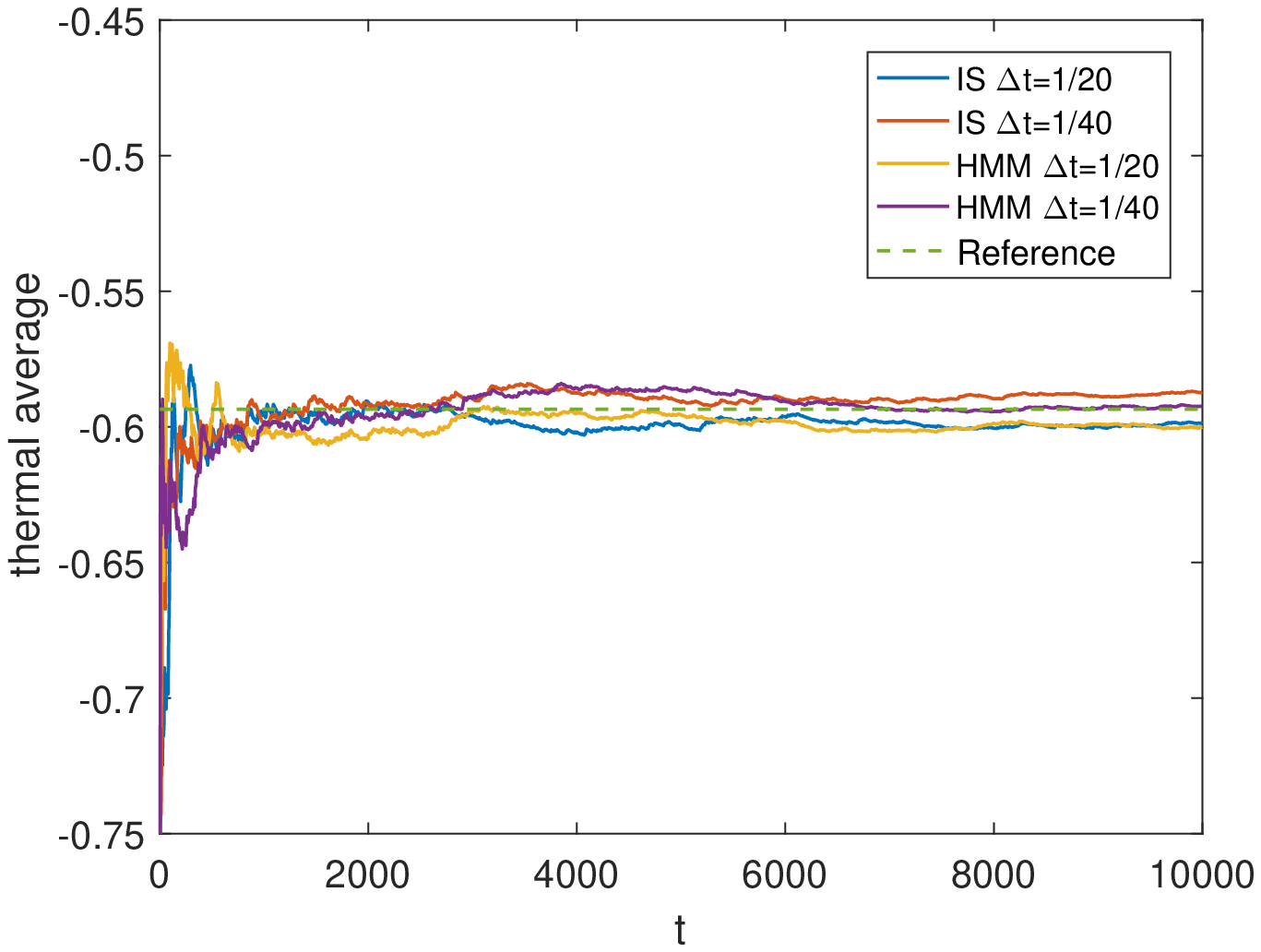}\\
\caption{ IS method and HMM method with $N= 4$, $\beta=1$ and various $\Delta t$ with potential model \eqref{ex:pot1}, $R=40$ in the HMM method.  Top: diagonal observable, the reference value is $0.850241$. Bottom: off-diagonal observable, the reference value is $-0.593497$.}
\label{fig:ISdt}
\end{center}
\end{figure}

In Figure~\ref{fig:ISdt}, we plot the numerical results with the IS
method and with the HMM method. We observe that for both observables
\eqref{eq:ob1} and \eqref{eq:ob2}, we are able to correctly
approximate the thermal averages for various time step sizes
$\Delta t$. For the HMM method, we choose a sufficiently large ratio
of the macro/micro solver step sizes, $R={40}$, and we observe very
similar results to those obtained by the IS method. This manifests a
major improvement in the PIMD method for sampling off-diagonal
observables.

\begin{figure}[htbp]
\begin{center}
\includegraphics[scale=0.55]{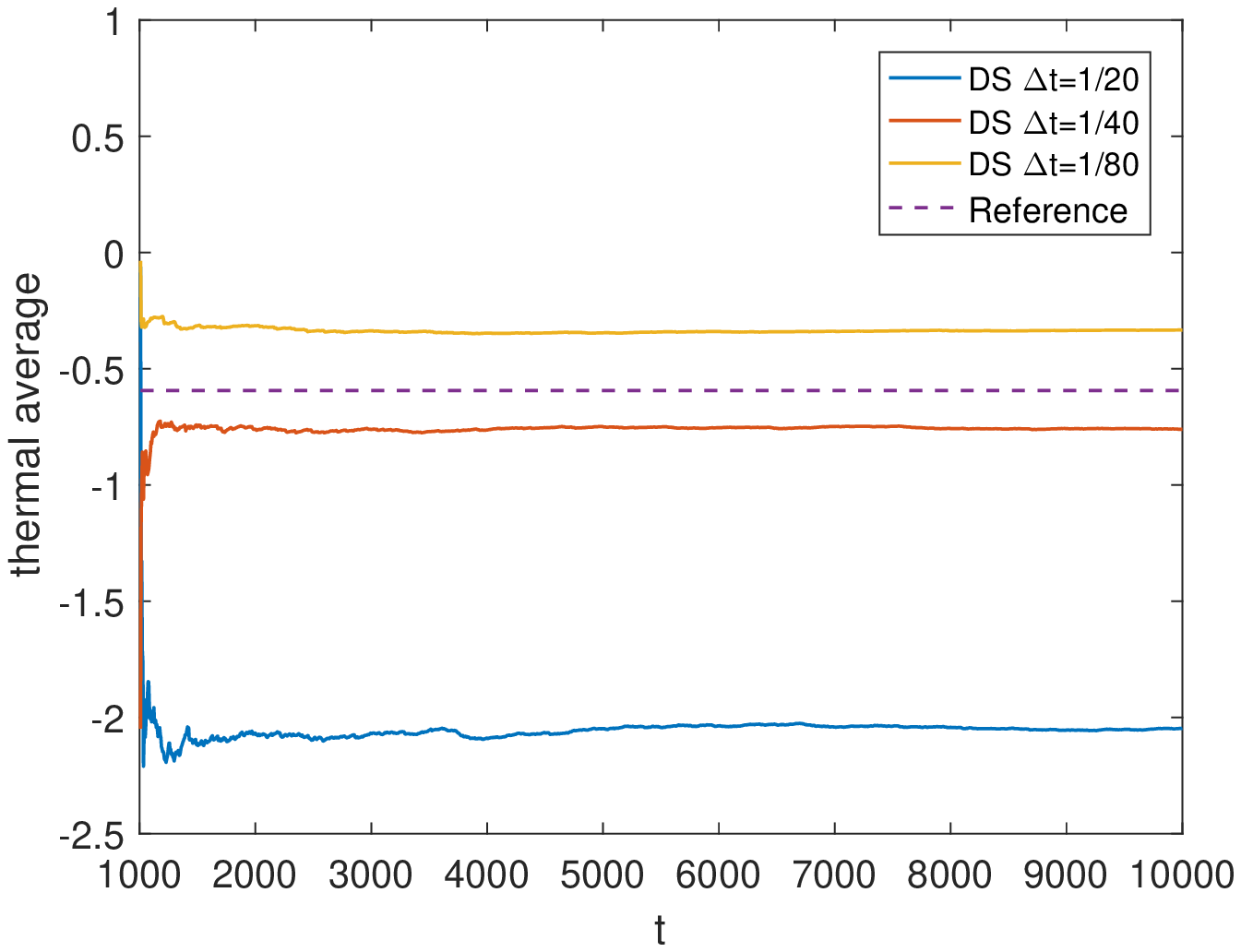} \includegraphics[scale=0.55]{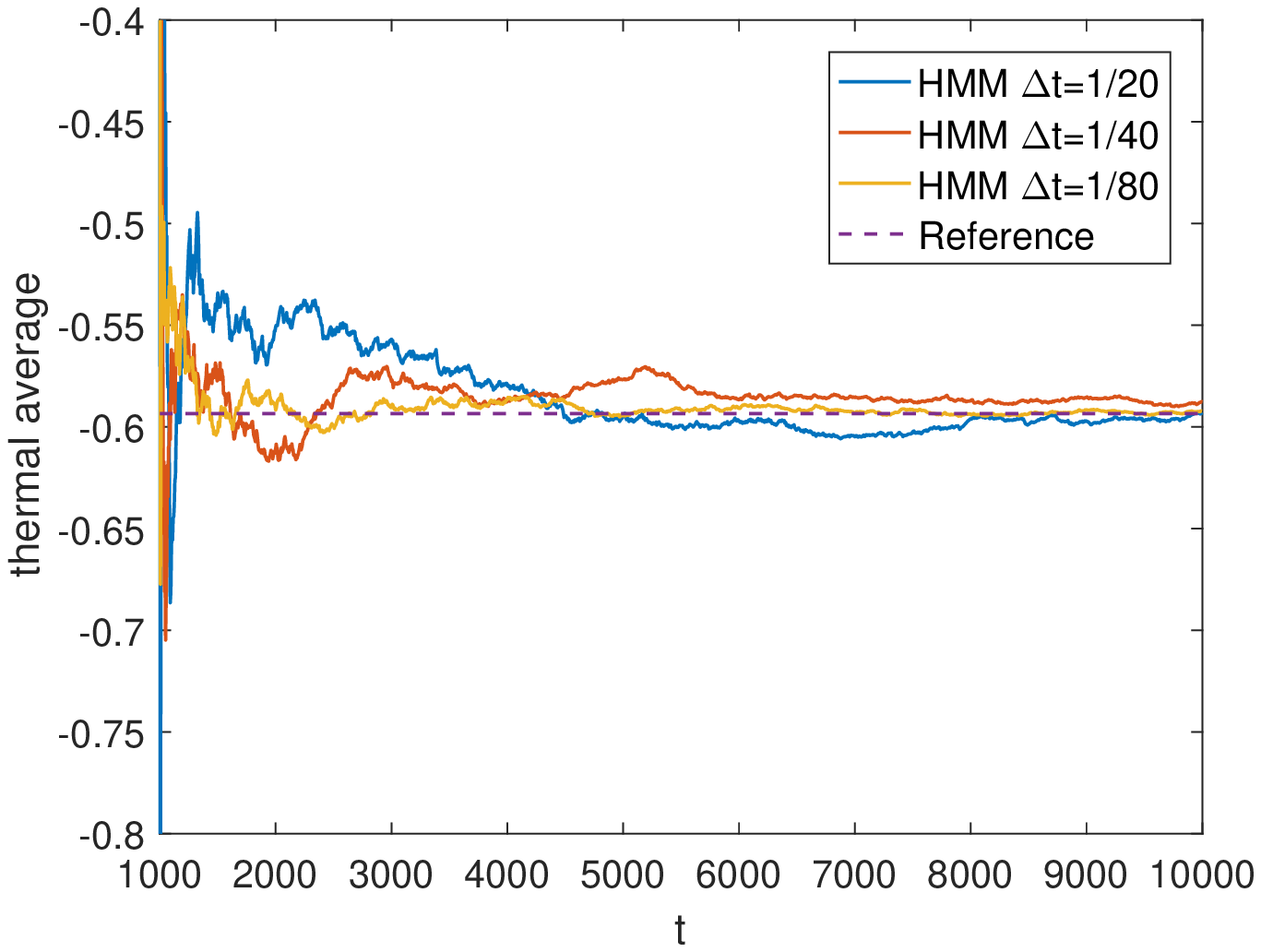}\\
\caption{ Off-diagonal observable  with potential model  \eqref{ex:pot1}. Top: DS method with $N=16$ and various $\Delta t$. Bottom: HMM method with the number of beads $N= 16$ and various $\Delta t$. The reference value is $-0.593497$.}
\label{fig:N16dt}
\end{center}
\end{figure}

We repeat the numerical tests for the off-diagonal observable
\eqref{eq:ob2}, for a larger number of beads, $N= 16$, $\beta=1$ and
various time step sizes $\Delta t$, with the DS method and the HMM
method. The tests for the diagonal observable are skipped since it is
less challenging, and similar to the case when $N=4$, the DS method
can already do a good job in those tests.  In the HMM method, we
choose the ratio of the macro/micro solver step sizes $R={40}$. The numerical results
are plotted in Figure \ref{fig:N16dt}, where we observe that when $N$
is larger, the numerical errors of the DS method become also larger
with same $\Delta t$. On the other hand, the HMM method
give accurate approximation even for fairly large $\Delta t$.

{
We remark that, these two tests above are presented in order to directly compare with the numerical results in \cite{LuZhouPIMDSH} and show the improvements. To rule out the possibility that the DS method only works for simple diagonal observables, we show in the following the test with another diagonal observable
\begin{equation} \label{eq:ob3}
\wh A=\begin{bmatrix}
  e^{- {\wh q}^2}   &  0\\
 0 &   -e^{- {\wh q}^2}  
\end{bmatrix},
\end{equation}
where the sign of the observable changes on different diabatic
surfaces. We still use the potential model \eqref{ex:pot1}, and choose
$N=4$, $\beta=1$. We test the DS method with
$\Delta t = \frac{1}{20}$, $\frac{1}{40}$, $\frac{1}{80}$ and
$\frac{1}{160}$, which are compared the IS method and the HMM method
(with $R=40$) for $\Delta t = \frac{1}{20}$ and $\frac{1}{40}$, and
the results are plotted in Figure~\ref{fig:diag2}. We observe that,
unlike the tests with off-diagonal observables, the three methods give
consistent numerical results even for large $\Delta t$, but the IS
method and the HMM method give superior performances to the DS method.
\begin{figure}[htbp]
\begin{center}
\includegraphics[scale=0.55]{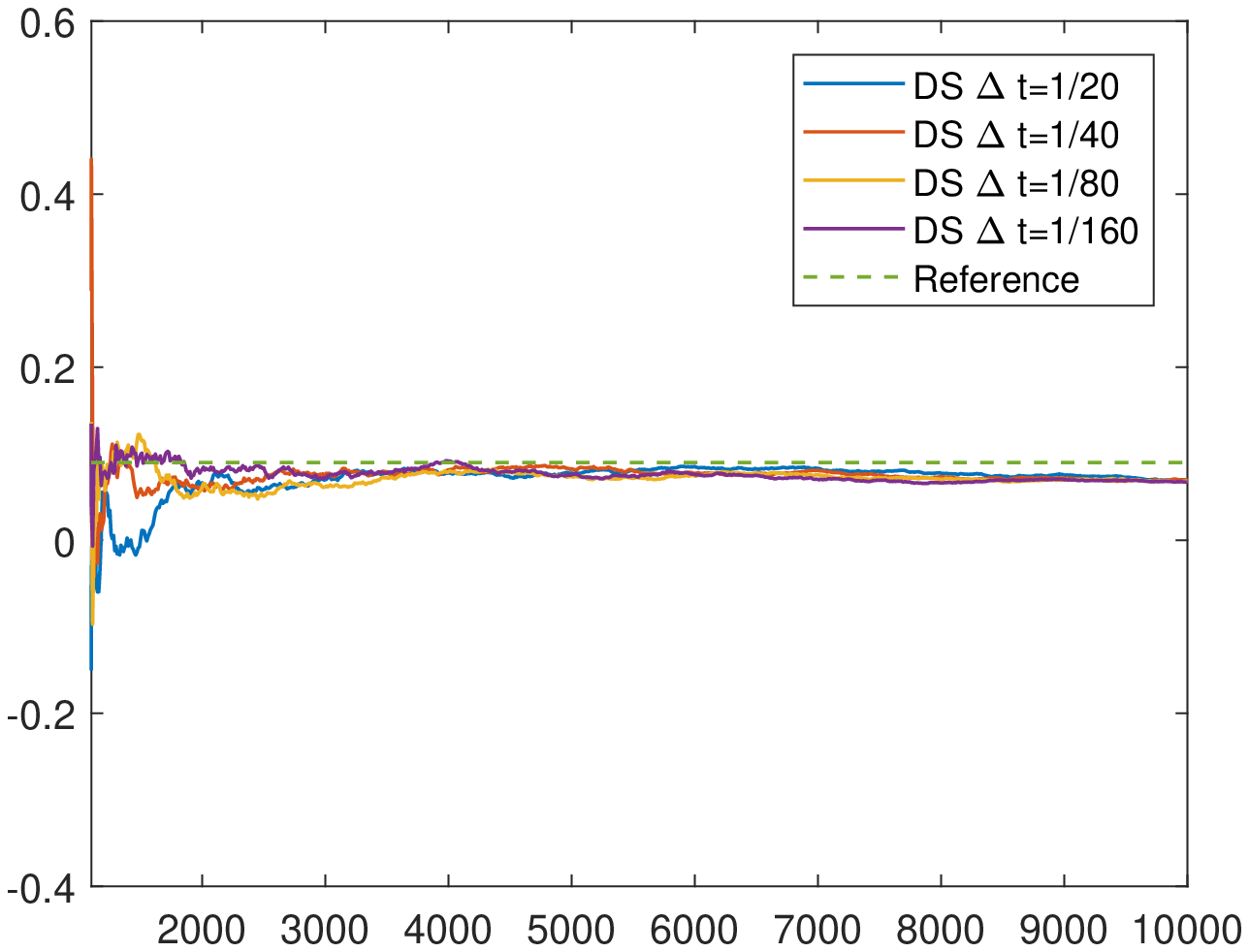} \includegraphics[scale=0.55]{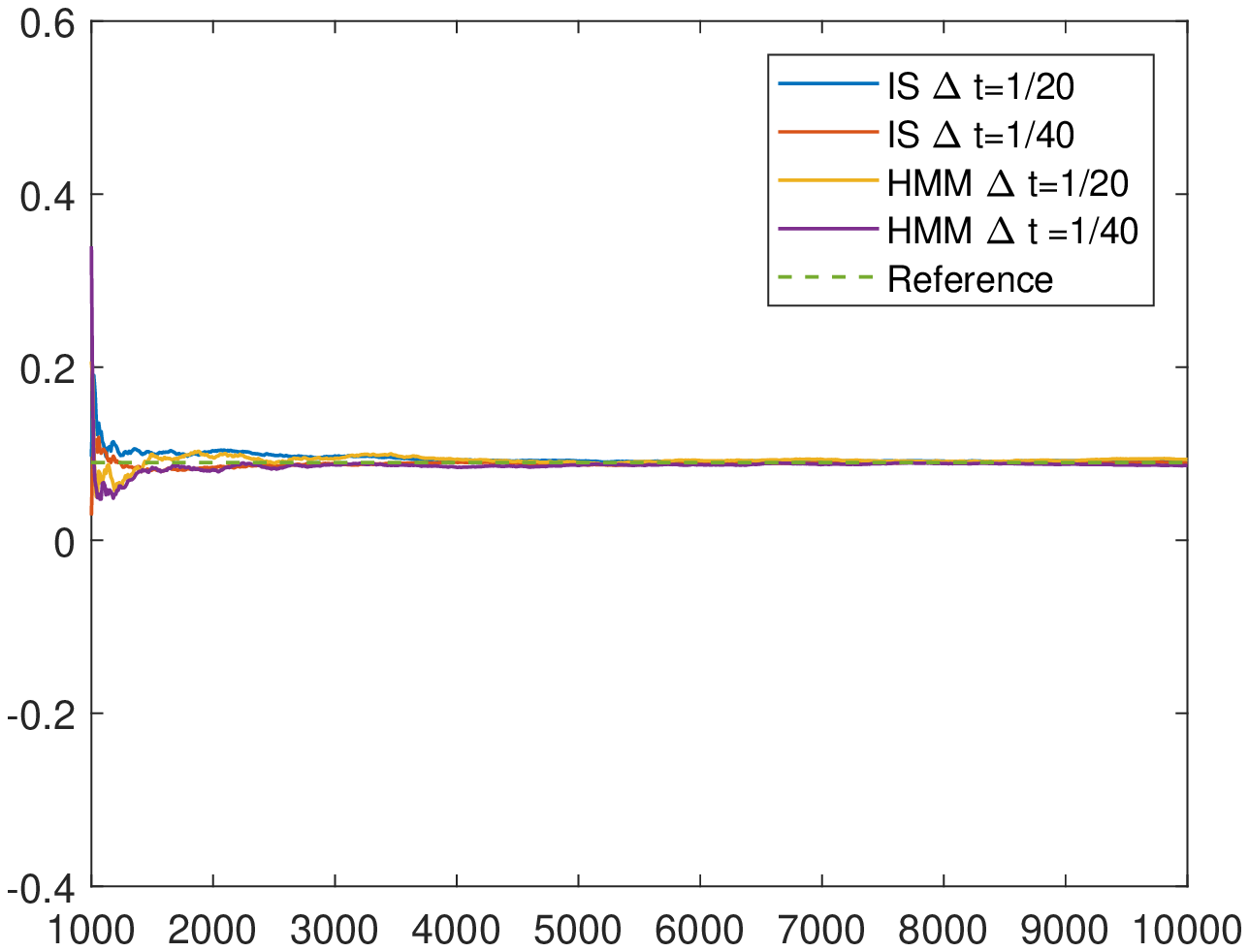}\\
\caption{ Diagonal observable \eqref{eq:ob3} with potential model  \eqref{ex:pot1}. Top: DS method with various $\Delta t$. Bottom: IS method and HMM method with various $\Delta t$. The reference value is $0.089901$. }
\label{fig:diag2}
\end{center}
\end{figure}
}

\begin{figure}[htbp]
\begin{center}
\includegraphics[scale=0.55]{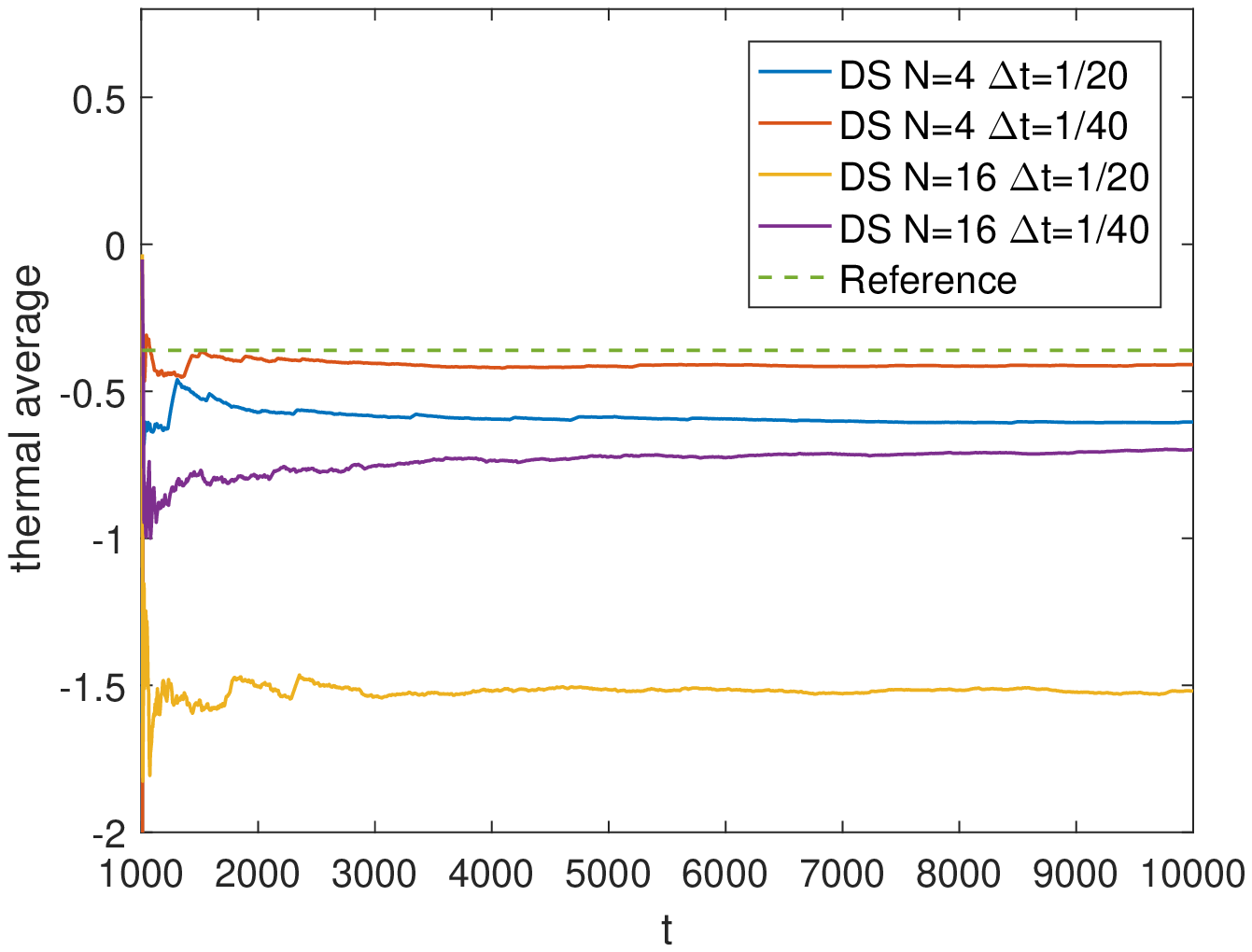} \includegraphics[scale=0.55]{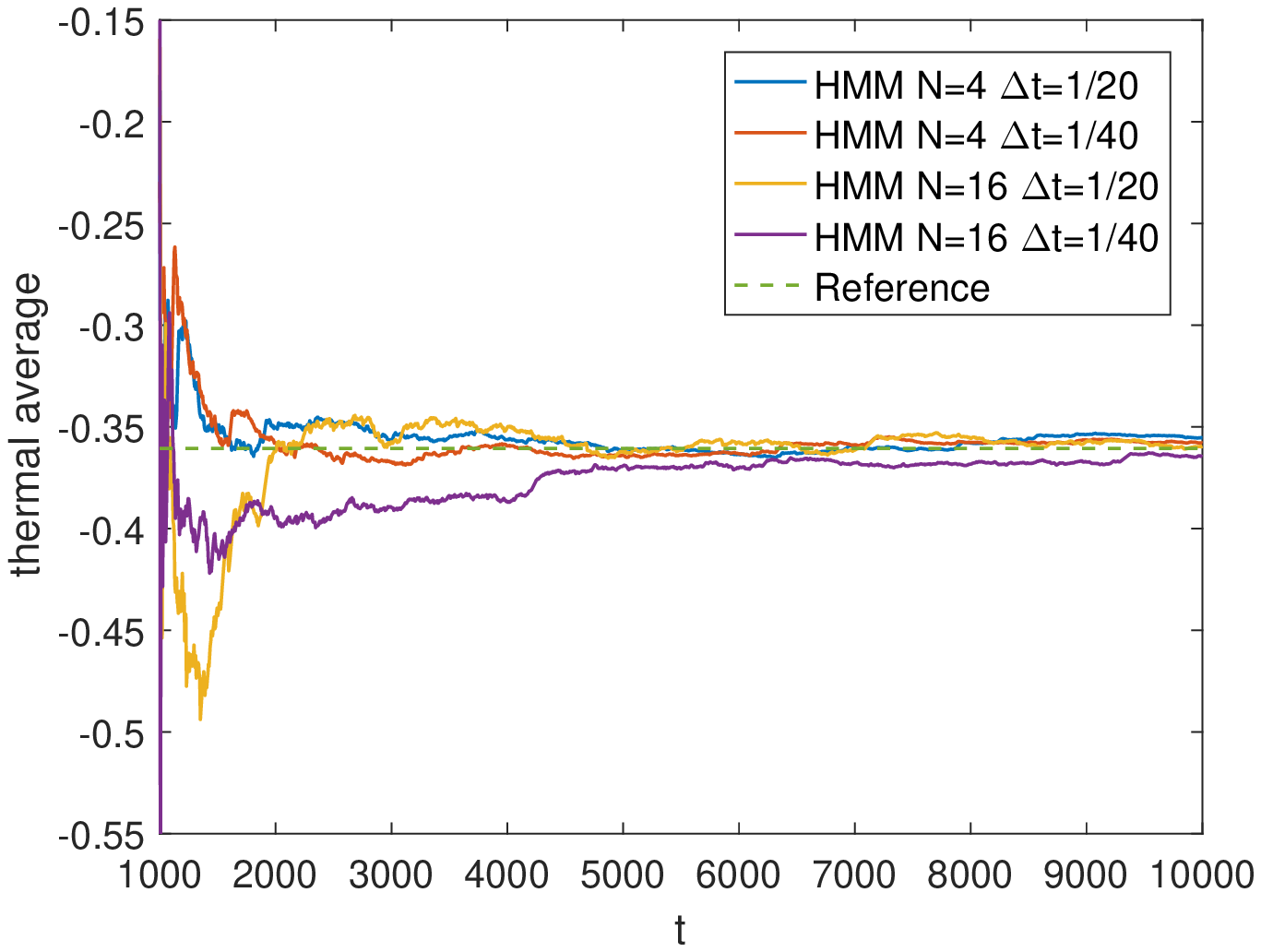}\\
\caption{ Off-diagonal observable with potential model  \eqref{ex:pot2}. Top: DS method with $N=16$ and various $\Delta t$. Bottom: HMM method with the number of beads $N= 16$ and various $\Delta t$. The reference value is $-0.360548$.}
\label{fig:v2dt}
\end{center}
\end{figure}

Finally, we test with the second potential model \eqref{ex:pot2} for
the off-diagonal observable
\eqref{eq:ob2}. % which is asymmetric and the most active hopping zone
                % is not located at either minimum of the energy
                % surfaces.
The tests for the diagonal observable are skipped since it is less
challenging, and similar to the tests with potential model
\eqref{ex:pot1}, the DS method can already do a good job in those
tests. For $N=4$ and $N=16$, we implement the HMM method with $R=40$,
$\Delta t = \frac{1}{20}$ and $\frac{1}{40}$, which are compared with
the DS method. The numerical results are plotted in
Figure~\ref{fig:v2dt}, where we observe that, the HMM method gives
accurate approximation even for fairly large $\Delta t$, while the DS
method gives rather poor accuracy.

%In Figure~\ref{fig:HMMdt}, we plot the results with the PIMD with  multiscale integrator method. We choose a sufficiently large ratio of the macro/micro solver step sizes, $N_0=\frac{1}{40}$, and we basically observe very similar results to the  those obtained by the PIMD with infinite swap method.
%%\begin{figure}[htbp]
%\begin{center}
%\includegraphics[scale=0.5]{HMM_diag_dt.eps} \includegraphics[scale=0.5]{HMM_off_dt.eps}\\
%\caption{ PIMD with multiscale integrator method with $\beta_N= \frac 1 4$, $N_0=40$ and various $\Delta t$.  Left: diagonal observable. Right: off-diagonal observable.}
%\label{fig:HMMdt}
%\end{center}
%\end{figure}

%\subsection{convergence tests with different $\beta$?}

\subsection{Tests with different macro/micro time step ratios $R$}

In the following, we focus on further exploration of numerical
properties of the HMM method. We aim to test the sensitivity of the
HMM method on two parameters, the number of beads $N$ and the
macro/micro time step ratio $R$. In this section, we fix $N$ and test
with various $R$, and we present the tests with various $N$ with a
reasonably large $R$ in the next section.

\begin{figure}[htpb]
\begin{center}
\includegraphics[scale=0.55]{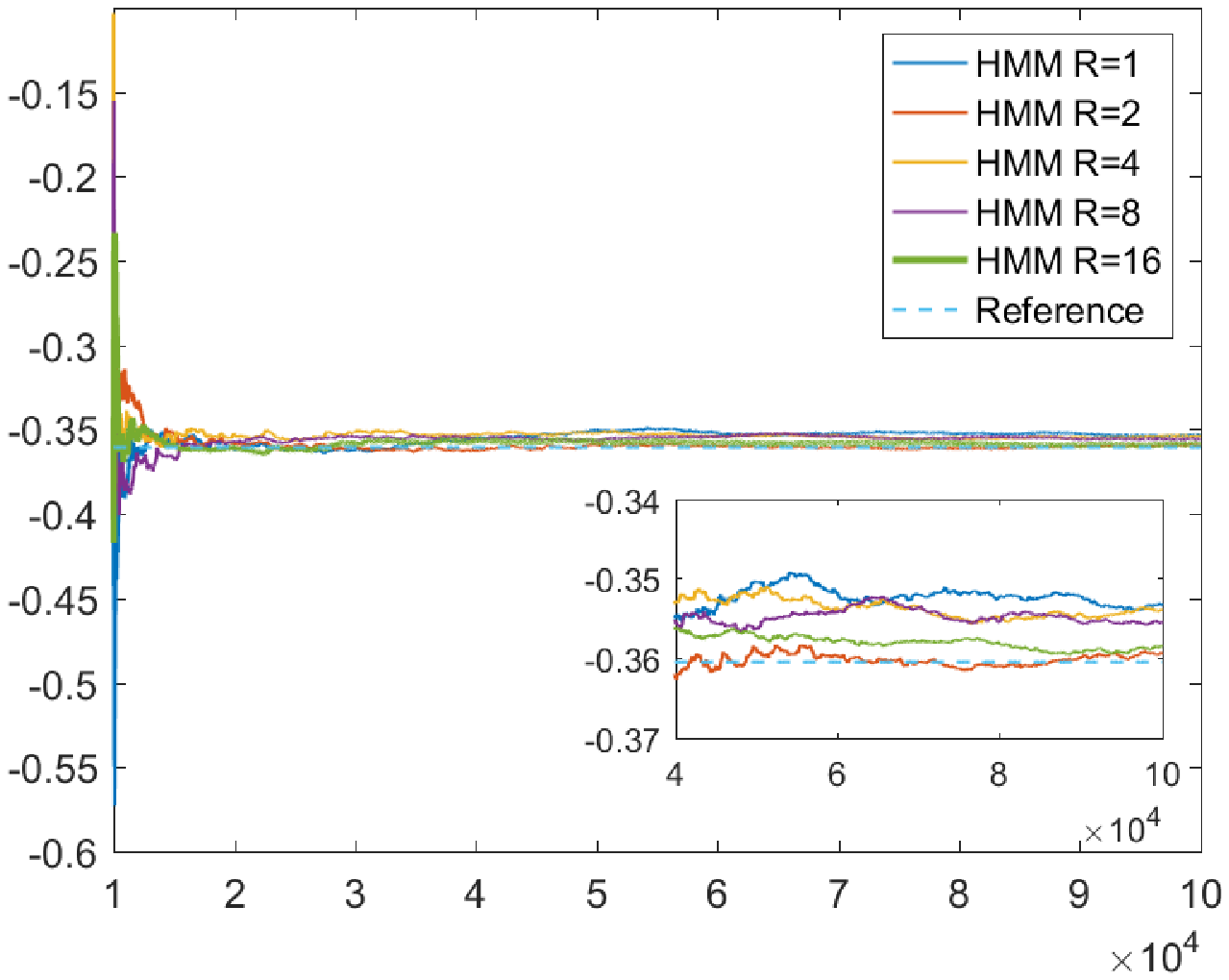} \\
\includegraphics[scale=0.55]{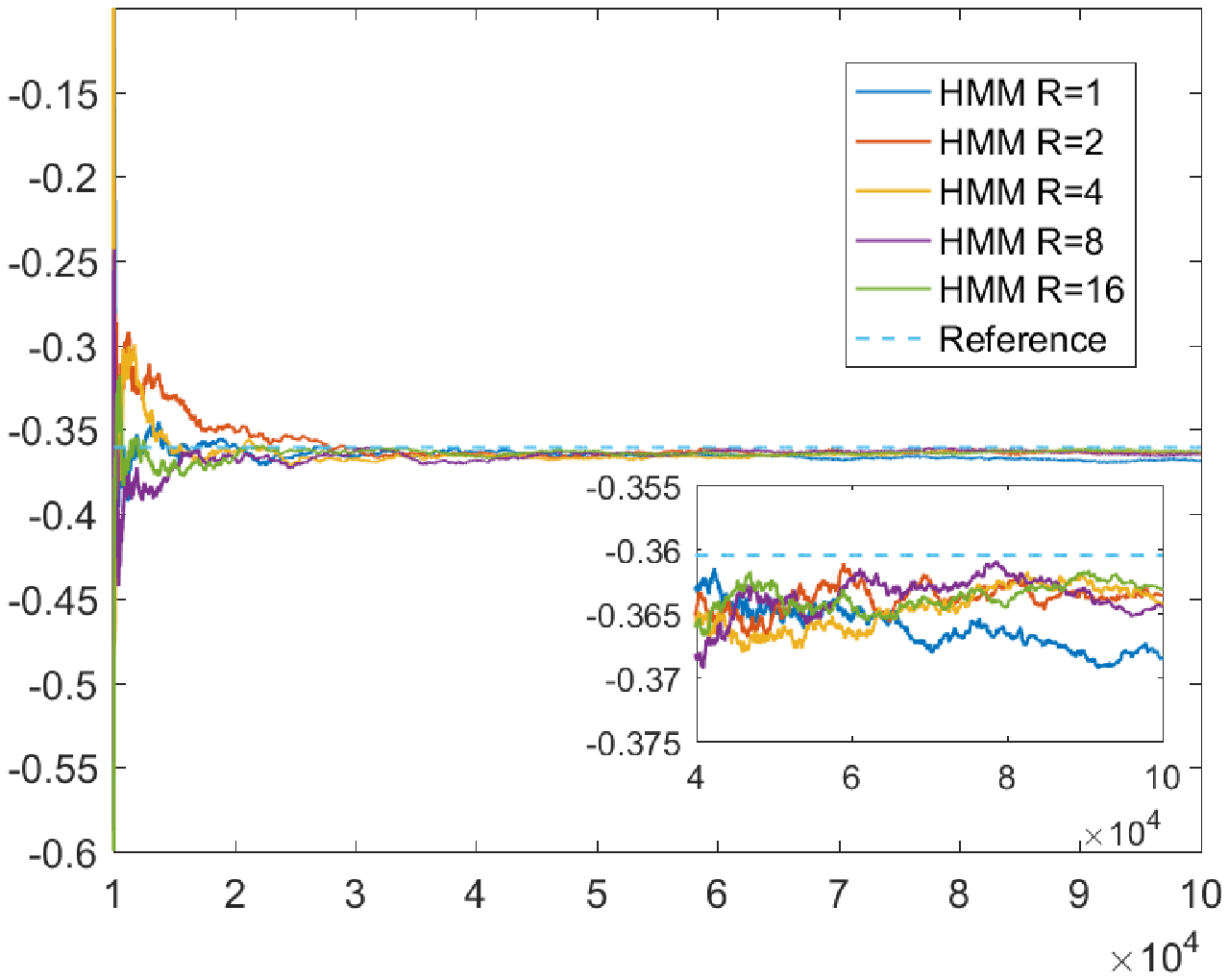} \\
\includegraphics[scale=0.55]{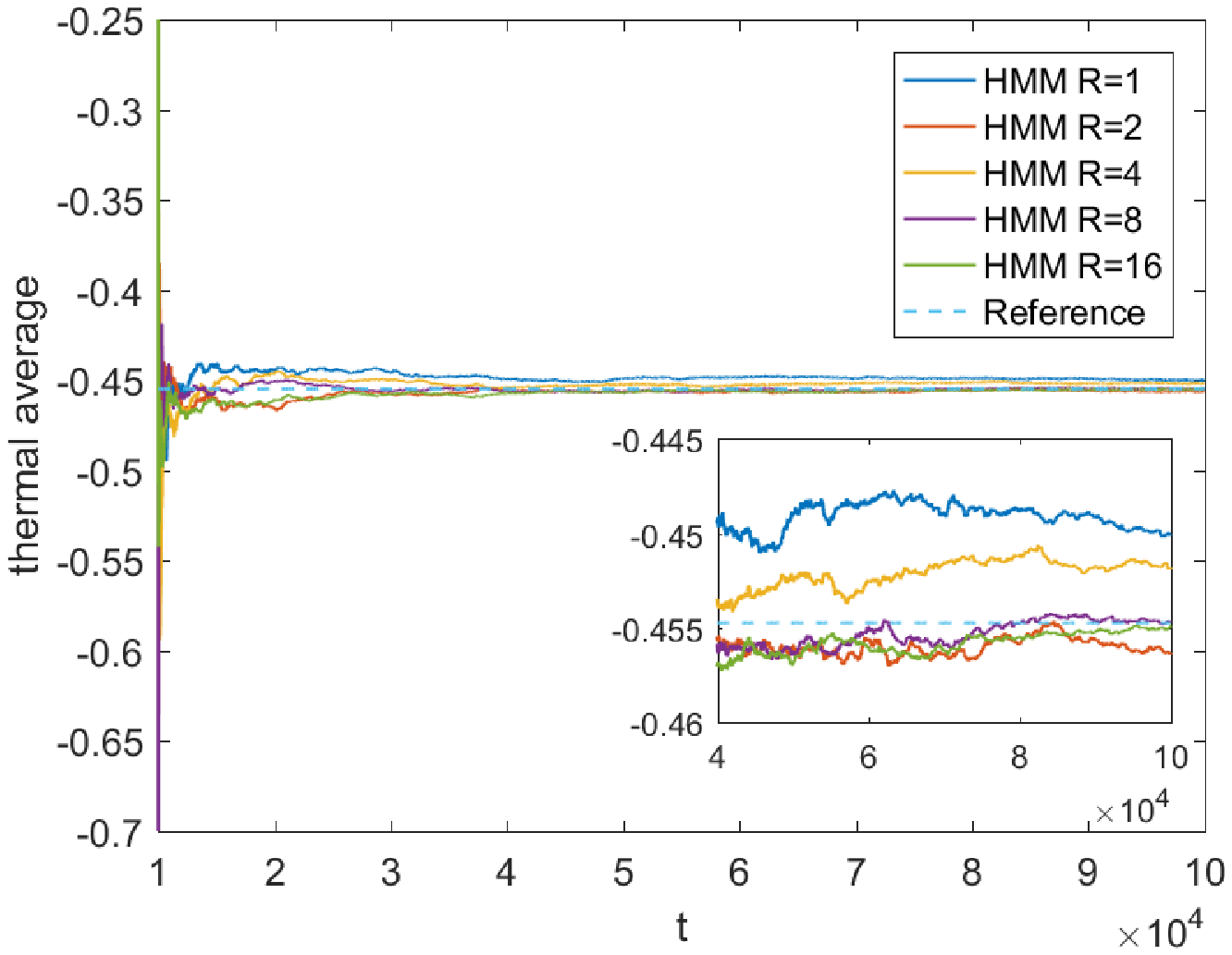}
\caption{The HMM method for potential \eqref{ex:pot2} with $T=10^6$,  $\Delta t=\frac {1}{20}$ and various $R$ for the off-diagonal observable. Top: 
$\beta=1$, $N=4$, the reference values is $-0.360548$. Middle: $\beta=1$, $N=16$,  the reference values is $-0.360548$. Bottom: $\beta=2$, $N=16$, the reference value is $ -0.454771$.}
\label{fig:diffN0}
\end{center}
\end{figure}

In this section we aim to test the HMM method with different $R$ with
a fairly large $\Delta t$, because the tests are obviously less
challenging if $\Delta t$ is smaller.  We will focus on the second
potential model \eqref{ex:pot2} as it is more challenging. We first
fix $N=4$, $\beta=1$ and $\Delta t=\frac {1}{20}$, and test with the
off-diagonal observable \eqref{eq:ob2}. We plot the results with
$R=1$, $2$, $4$, $8$ and $16$, which is also compared with the
reference value. We observe in Fig~\ref{fig:diffN0} (top) that even
when $R=1$, the HMM method behaves much better than the DS method, and
when $R=8$ and $16$, the HMM method behaves similarly, which suggest
good convergence with respect to the choice of $R$.

We repeat the tests for $\beta=1$, $N=16$ and for $\beta=2$, $N=16$,
where we observe similar trends as plotted in Fig~\ref{fig:diffN0}
(middle and bottom). We conclude that when the IS method is no longer
affordable since it has too many discrete states to explore, the HMM
method with a reasonably large $R$ provides an alternative much faster
way to sample off-diagonal observables.

%\begin{figure}[htbp]
%\begin{center}
%\includegraphics[scale=0.55]{N4diffN0.eps} \\
%\caption{ PIMD with the multiscale integrator method with $\beta_N= \frac 1 4$, $\Delta t=\frac {1}{20}$ and various $N_0$.  Results are computed the off-diagonal observable, which is compared with PIMD with infinite swap method.}
%\label{fig:HMMISdt}
%\end{center}
%\end{figure}

At last, we carry out another comparison test between the DS method
and the HMM method.  We choose $N=4$, $\beta=1$, the tests are
implemented with the second potential model \eqref{ex:pot2} for the
off-diagonal observable \eqref{eq:ob2}. In the DS method, we test with
$\Delta t = \frac{1}{20}$, $\frac{1}{40}$, $\frac{1}{80}$ and
$\frac{1}{160}$ till $T=10^6$. And the HMM method, we fixed
$\Delta t = \frac{1}{20}$, but $R=1$, $2$, $4$ and $8$, so that
$\Delta t/R$ in the HMM method matches $\Delta t$ in the DS
method. This test is fair in terms of the numerical hopping frequency
in both methods; we shall keep in mind though that the DS method takes
finer time steps in integrating the trajectory of $(\bd q, \bd p)$, so
it is more expensive.  The mean squared errors of the empirical
averages are defined as $\text{M.S.E.} = \text{Bias}^2 + \text{Var}$,
where $\text{Bias}$ is calculated using the reference value and
$\text{Var}$ is estimated using the observed data and the effective
sampling size.  The $\text{M.S.E.}$ for both methods are shown in
Table~\ref{table:testR}, where we clearly observe that the
performances of the HMM method are much better than the DS method. 
% We remark that, with the fixed ratio $\Delta t/R$, it is not clear
% yet how to choose the $dt$ and $R$ to optimized the accuracy, which
% we may explore in the future.

\begin{table}
  \centering
  \begin{tabular}{ c| c| c|c|c} \hline
    DS & $\Delta t=\frac{1}{20}$ & $\Delta t=\frac{1}{40}$ & $\Delta t=\frac{1}{80}$ & $\Delta t=\frac{1}{160}$   \\ \hline
M.S.E. & 6.5666e-2 & 2.6749e-3 & 9.7038e-4 & 8.4174e-4 
    \\ \hline
   HMM & $\frac{\Delta t}{R}= \frac{1}{20}$ & $\frac{\Delta t}{R}= \frac{1}{40}$ & $\frac{\Delta t}{R}= \frac{1}{80}$ & $\frac{\Delta t}{R}= \frac{1}{160}$
    \\ \hline
M.S.E. & 1.2851e-4 & 3.5028e-5 & 5.8980e-5 & 3.7058e-5   \\ \hline
\end{tabular}
\caption{Mean squared errors in the DS  method and in the HMM method. In the HMM method, we fix $\Delta t= \frac{1}{20}$, but $R$ varies. The reference value is $-0.360548$. }
  \label{table:testR}
\end{table}

\subsection{Tests with different $\beta_N$ for fixed $R$}

\begin{figure}[htbp]
\begin{center}
\includegraphics[scale=0.55]{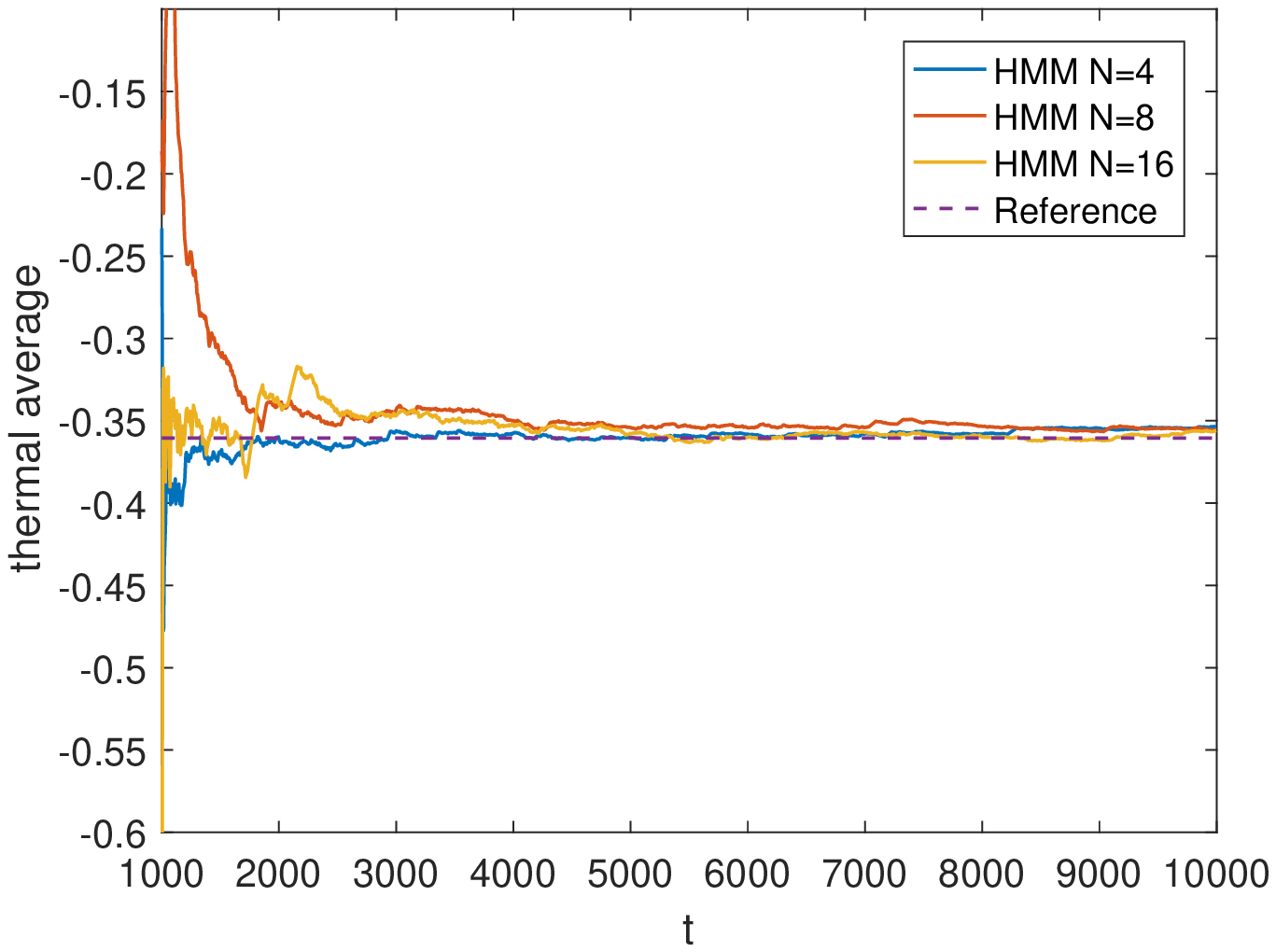}\\
 \includegraphics[scale=0.55]{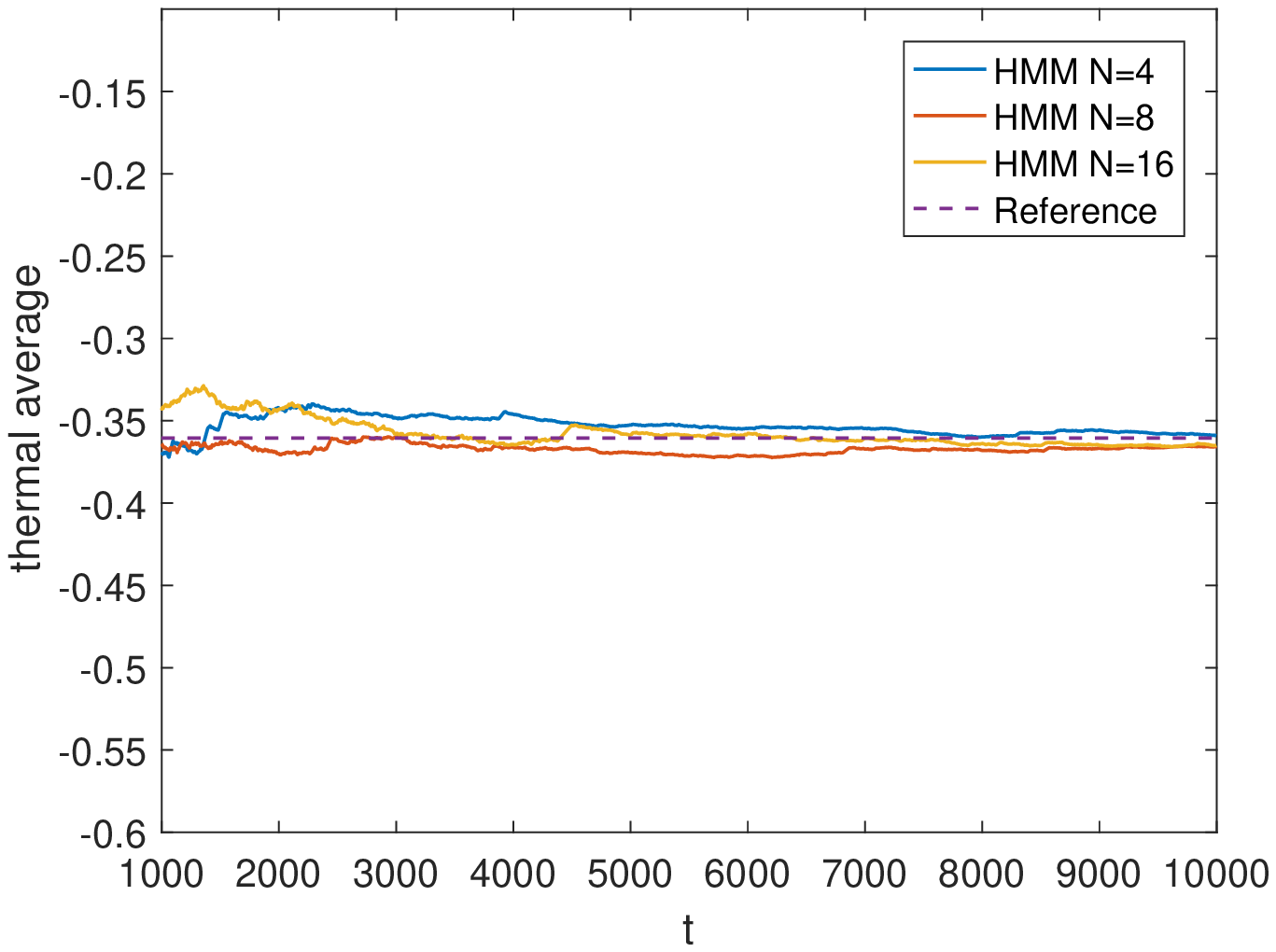}\\
 \includegraphics[scale=0.55]{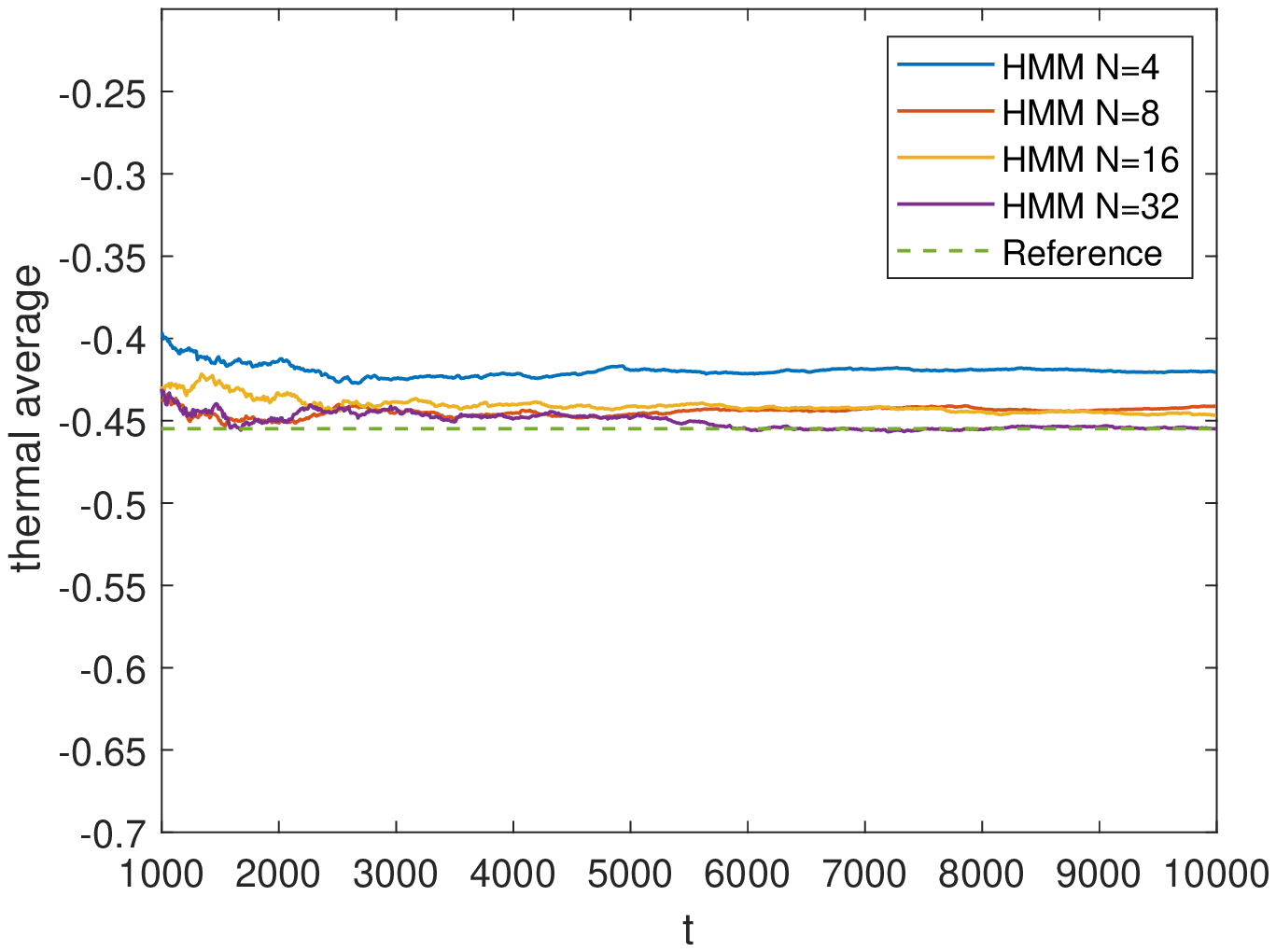}
 \caption{ {The HMM method for potential \eqref{ex:pot2} with
   $\Delta t= \frac{1}{50}$, $T=10^4$ and various choice of $\beta$,
   $R$ and $N$.  Top: $\beta=1$, $R=16$, the reference values is
   $-0.360548$. Middle: $\beta=1$, $R=32$, the reference values is
   $-0.360548$. Bottom: $\beta=2$, $R=32$, the reference value is
   $ -0.454771$.}}
\label{fig:diffN}
\end{center}
\end{figure}

In this section we aim to test the HMM method with various $N$ for a
fixed but reasonably large $R$. We will again stick to the second
potential model \eqref{ex:pot2} with the off-diagonal observable
\eqref{eq:ob2}.  We fixed $\Delta t=
\frac{1}{50}$, %\jl{$dt$ or $\Delta t$?}
the final time $T=10^4$, $\beta=1$ and $R=16$, and test the
performance of the HMM method with the number of beads $N=4$, $8$,
$16$. We test with $\beta=1$, $R=32$ and with $\beta=2$ and $R=32$,
and the results are plotted in Figure~\ref{fig:diffN} (middle and
bottom). We observe that, when $\beta=1$, for $R=16$ or $32$, the HMM
method already gives accurate results. However, when $\beta=2$ and
$R=32$, the numerical error is significantly larger for $N=4$. We
believe it is mainly due to the asymptotic error, since the errors are
reduced for $N=8$ and $N=16$, {and become negligible for $N=32$}.

Together with the previous section, we conclude that, the performance
of the HMM method is more sensitive on the number of beads $N$ than to
the time step ratio $R$. In practice, one needs to choose a proper $N$
to make sure the asymptotic error is acceptable and subsequently
choose a reasonably large $R$, which is less constrained. We would
like to remark that, choosing a sufficiently large $N$ might be a
challenge, which is shared by most PIMD methods. Roughly speaking,
larger $N$ is needed when the temperature is low ($\beta$ is large
accordingly) and when profiles of the energy surfaces are complicated.

%
%Besides, it seems to suggest that the approximation error introduced by a rough sample in the surface configuration space gives error which is independent of the number of beads.  

\section{Conclusion}

Based on the theoretical analysis of the infinite swapping limit of
the previously developed path-integral molecular dynamics with surface
hopping method, we proposed in this work a multiscale integrator for
the infinite swapping limit for sampling thermal equilibrium average
of multi-level quantum systems. The efficiency of the proposed
sampling scheme is greatly improved compared with the direct
simulation of the PIMD-SH method. As for the future works, one
immediate direction to explore is to combine the equilibrium sampling
scheme with real time surface hopping dynamics to calculate dynamical
correlation functions. {In addition, we plan to test and further
  improve the proposed algorithms for realistic chemical applications
  with multidimensional potential surfaces.}

\begin{acknowledgments}
  This work is partially supported by the National Science Foundation
  under grant DMS-1454939. The work of Zhennan Zhou is also partially
  supported by a start-up fund from Peking University.
\end{acknowledgments}

\appendix

{
\section{Ring polymer representation for  two-level quantum systems} \label{Append:A}

In this part, we present the details of the ring polymer representation of thermal averages as in \eqref{eq:aveA}  for two-level quantum systems, which have been rigorously derived in \cite{LuZhouPIMDSH}. With the diabatic basis, we approximate the partition
function by a ring polymer representation with $N$ beads
\begin{multline}
\label{eq:reZ} \tr_{ne}[e^{-\beta \wh H}] \approx
\mc{Z}_N  := \frac{1}{(2\pi)^{dN}} \int_{\RR^{2dN}} \ud \bd q \ud \bd p \times \\
\times \sum_{\bd{\ell} \in \{0, 1\}^N}  \exp(-\beta_N H_N(\bd{q},
\bd{p}, \bd{\ell})),
\end{multline} where $\beta_N=\beta/N$. The ring polymer that consists
of $N$ beads is prescribed by the configuration
$(\bd{q}, \bd{p}, \bd{\ell}) \in \RR^{dN} \times \RR^{dN} \times \{0,
1\}^N$.

For a given ring
polymer with configuration $(\bd{q}, \bd{p}, \bd{\ell})$, the
effective Hamiltonian $H_N(\bd{q}, \bd{p}, \bd{\ell})$ is given by
\begin{equation}\label{eq:Ham} H_N(\bd q, \bd p, \bd\ell) =
\sum_{k=1}^N \bra{\ell_k} G_k \ket{\ell_{k+1}},
\end{equation} 
where we take the convention that $\ell_{N+1} = \ell_1$
and matrix elements of $G_k$, $k = 1, \ldots, N$, are given by
\begin{subequations} \label{eq:Gele}
\begin{multline}
  \bra{\ell} G_{k} \ket{\ell'} = \frac{p^2_k}{2M}+ \frac{ M\left(q_k-q_{k+1}
\right)^2 }{2(\beta_N)^2} \\
 +\frac{V_{00}(q_k)+V_{11}(q_k)}{2}
-\frac{1}{\beta_N}\ln\Bigl( \sinh \bigl(\beta_N \Abs{V_{01}(q_k)}
\bigr) \Bigr), 
\end{multline}
for $\ell \ne \ell'$, and the diagonal terms are given as
\begin{multline} \bra{\ell} G_{k} \ket{\ell} = \frac{p^2_k}{2M}+
  \frac{M \left(q_k-q_{k+1} \right)^2}{2(\beta_N)^2} \\
  +V_{\ell\ell}(q_k) - \frac{1}{\beta_N} \ln \Bigl( \cosh \bigl(
  \beta_N \Abs{ V_{01} (q_k) }\bigr) \Bigr),
\end{multline}
\end{subequations}
where we have suppressed the $\bd{q}$ and $\bd{p}$ dependence in the
notation of $G_k$.  Here $G_k$ can be understood as the the contribution of $\bra{q_k} e^{-\beta_N \wh H} \ket{q_{k+1}}$
to the effective Hamiltonian $H_N$ in the ring polymer representation. The readers may refer to \cite{LuZhouPIMDSH} for the derivations.

For an observable $\wh{A}$, under the ring polymer representation, we
have
\begin{multline} \label{eq:reAo} \tr_{ne}[e^{-\beta \wh H}\wh A ]
\approx \frac{1}{(2\pi)^{dN}} \int_{\RR^{2dN}} \ud \bd q \ud \bd p
\sum_{\bd{l} \in \{0, 1\}^N} \\ \times \exp(-\beta_N H_N) W_N[A],
\end{multline} where the weight function associated to the observable
is given by (recall that $\wh{A}$ only depends on position by our
assumption)
\begin{multline}\label{eq:WNA} W_N[A] (\bd q, \bd p, \bd\ell) =
  \frac{1}{N} \sum_{k=1}^N \biggl( \bra{\ell_k} A(q_k) \ket{\ell_k} \\
  - e^{\beta_N \langle \ell_k| G_k| \ell_{k+1} \rangle - \beta_N
    \bra{\bar{\ell}_k} G_{k} \ket{ \ell_{k+1}}} \bra{\ell_{k}}
  A(q_{k}) \ket{\bar{\ell}_k}
  \frac{V_{\ell_k\bar{\ell}_k}}{\abs{V_{\ell_k\bar{\ell}_k}}} \biggr),
\end{multline}
where we have introduced the short hand notation
$\bar{\ell}_k = 1 - \ell_k$, \textit{i.e.}, $\bar{\ell}_k$ is the
level index of the other potential energy surface than the one
corresponds to $\ell_k$ in our two-level case.  Similar as for the
partition function, the ring polymer representation \eqref{eq:reAo}
replaces the quantum thermal average by an average over ring polymer
configurations on the extended phase space
$\R^{dN} \times \R^{dN} \times \{ 0, 1 \}^N$.

The ring polymer representation for a multi-level quantum system can
be also constructed using the adiabatic basis
\cite{SchmidtTully2007,LuZhouPIMDSH}, and much of the current work
also extends to the ring polymer with the adiabatic basis. We will skip
the details and leave to interested readers.
}

\bibliography{surfacehoppingPIMDIS} 
\end{document}